\documentclass[fleqn,usenatbib]{mnras}

\usepackage{prettyref}
\usepackage{float}
\usepackage{rotfloat}
\usepackage{amsmath}
\usepackage{amssymb}
\usepackage{graphicx}
\usepackage{color}

\usepackage[T1]{fontenc}
\usepackage{ae,aecompl}

\usepackage{tablefootnote}

\makeatletter

\usepackage{indentfirst}
\usepackage{leftidx}

\makeatother

\usepackage[normalem]{ulem}
\usepackage{tikz}
\usetikzlibrary{shapes,arrows}
\usepackage{amsmath}
\usepackage{amssymb}
\usepackage{wasysym}
\usepackage{tabularx}
\usepackage{array}
\newcommand{\vect}[1]{\boldsymbol{#1}}
\newcommand{\be}{\begin{eqnarray}}
\newcommand{\ee}{\end{eqnarray}}

\title[Orbital Decay in Binaries]{Orbital Decay in Binaries Containing Post-Main Sequence Stars}
	
\author[M. Sun et al.]{M. Sun$^{1}$\thanks{E-mail: msun@virginia.edu}, P. Arras$^{1}$, N. N. Weinberg$^{2}$, N. W. Troup$^{1}$, and S. R. Majewski$^{1}$
\\	
$^{1}${Department of Astronomy, University of Virginia, P.O. Box 400325, Charlottesville, VA 22904, USA}\\
$^{2}${Department of Physics, and Kavli Institute for Astrophysics and Space Research,} \\
  {Massachusetts Institute of Technology, Cambridge, MA 02139, USA}}

\date{Accepted 2018 September 6. Received 2018 August 4; in original form 2018 February 18}

\pubyear{2018}
	
\begin{document}
\label{firstpage}
\pagerange{\pageref{firstpage}--\pageref{lastpage}}
\maketitle
	
\begin{abstract}

The orbital decay of binaries containing a primary sub-giant or red giant star and a stellar or substellar companion is investigated. The tide raised in the primary by the companion leads to an exchange of angular momentum between the orbit and the stellar spin, causing the orbit to contract and the primary to spin up. The rate of orbital decay is computed including both the equilibrium tide, damped by turbulent viscosity in the convective envelope, and the dynamical tide, assumed to be a traveling internal-gravity wave in the radiative core. For close binaries, the tidal forcing period is expected to be much shorter than the eddy turnover timescale in the convective envelope, and the prescription for ``reduced" viscosity is an important consideration. The dynamical tide tends to dominate for the closest orbits, while the equilibrium tide dominates for more distant orbits, with the crossover point depending on the stellar mass. The spin up of the primary to synchronous rotation occurs for sufficiently massive secondaries, and this greatly slows the orbital decay until the Darwin instability occurs. A parameter survey is presented for orbital decay as a function of primary and secondary mass, as well as turbulent viscosity prescription. These results are summarized with analytic formulae and numerical results for the age-dependent critical separation, $a_{\rm crit}$, inside of which orbital decay is rapid, and few systems are expected to be observed. The calculations of $a_{\rm crit}$ are compared with APOGEE binaries, as well as solar mass exoplanet host stars.

\end{abstract}
	
	
\section{Introduction}
Tidal friction becomes orders of magnitude larger as stars leave the main sequence (MS) and ascend the red giant branch (RGB). Binaries that suffered relatively weak tidal effects on the MS may suffer catastrophic orbital decay during the sub-giant branch (SGB) \citep{2013ApJ...772..143S}, RGB and asymptotic giant branch, resulting in the binary components coming into contact. Sufficiently massive secondaries, $M_2 \ga 10^{-2}\, M_\odot$, may then initiate a common envelope spiral-in and ejection of the primary's envelope \citep{1976IAUS...73...75P}, forming a close binary containing the secondary and a helium-core white dwarf. Smaller secondaries may give rise to mergers with the helium core of the primary, or destruction in the envelope of the primary \citep{1998A&A...335L..85N}.
	
The engulfment of companions proceeds from smaller to larger orbital separation. Most previous studies (e.g. \citealt{2008MNRAS.386..155S,2011ApJ...737...66K,2012ApJ...761..121M,2014ApJ...794....3V}) have focused on the end result, namely the critical separation outside of which systems may survive as the primary transitions into a white dwarf. This paper focuses on earlier stages, when the primary has left the MS but is well below the tip of the RGB, as these are more commonly found in spectroscopic surveys. 
			
This paper was motivated by the close binaries found in the APOGEE survey. Employing three years of APOGEE \citep{2017AJ....154...94M} observations from the twelfth data release of the Sloan Digital Sky Survey \citep{2015ApJS..219...12A}, \cite{2016AJ....151...85T} compiled a catalog of 376 newly-found close binary systems. These are single-lined binaries with radial velocity fits for the orbit and secondary minimum mass, and a large range of Galactocentric radius, metallicity and evolutionary state of the primary. This sample is unique, as it contains both dwarf and giant primaries, and secondaries ranging from planetary to stellar masses. The present paper focuses on the 180 primaries which are post-MS, as theory predicts these systems have much stronger tidal friction. Interestingly, brown dwarf (BD) secondaries are nearly as common as stellar-mass secondaries in the APOGEE-1 sample. The putative ``brown-dwarf desert", a lack of close binaries with solar-type primaries and BD secondaries, is {\it not} found in this sample, in seeming contradiction to decades of previous surveys of FGK dwarfs \citep{2006ApJ...640.1051G}. The presence of secondaries, from planetary to brown dwarf to stellar mass, allows a test of synchronization and orbital decay over the entire range, from small-secondary Darwin unstable systems, which will come into contact due to orbital decay, as well as large-secondary systems, which will quickly synchronize. The latter then evolve on the RGB evolutionary time of the primary, with the spin frequency nearly equal to the orbital frequency as the star evolves.

Both equilibrium and dynamical tides are considered in this paper. For the equilibrium tide, shearing of the tidally forced fluid motion is dissipated as heat by turbulent viscosity in the convective envelope. As gravity waves are evanescent there, and the frequencies are far below acoustic waves, the fluid motion is not wavelike. Rather, fluid nearly follows equipotential surfaces, and large scale shearing motions are present. This fluid motion is sometimes approximated with the analytic ``equilibrium tide" solution out of convenience, although that derivation is only formally valid in radiative zones \citep{1998ApJ...502..788T}. However, since the surface of the star is nearly an equipotential, this analytic solution performs well, and is convenient. It is used in this paper. The fluid motions in the convective zone are damped by turbulent viscosity, in which resonant turbulent eddies due to thermal convection transport momentum and damp the tidal shear flow. The dissipation rate for this process depends on the uncertain details of the interaction of small-scale turbulence with a mean flow, but the two theories discussed in this paper have scalings $P_{\rm orb}^{-4}$ and $P_{\rm orb}^{-5}$ for the dissipation rate, and hence decrease much more slowly than the dynamical tides' $P_{\rm orb}^{-7.7}$. Hence it is expected that the dynamical tide dominates at small separation and the equilibrium tide at larger separation. One of the aims of this paper is to estimate the critical separation at which dynamical and equilibrium tide dissipation are comparable.

The dissipation of the equilibrium tide by turbulent viscosity in convection zones was first developed by \citet{1977A&A....57..383Z}. In that paper, a turbulent viscosity $\nu_\ell = v_\ell \ell / 3$ was proposed, where $\ell \sim H$  and $v_\ell \sim (F/\rho)^{1/3}$ are the size and velocity of the large, energy-bearing eddies. Here $H$ is the pressure scale height, $F$ is the heat flux, and $\rho$ is the mass density.
In many situations where this theory is applied, the large eddy turn-over time, $\tau_\ell = \ell / v_\ell$ is much longer than the forcing period, $P_{\rm f}$, and the eddies cannot efficiently transport momentum and damp the shear flow. \citet{1989A&A...220..112Z} proposed that large, non-resonant eddies move a small fraction $P_{\rm f}/2\tau_\ell$ of an overturn, and so the eddy turnover time should be reduced by this linear-in-period factor. \citeauthor{1977Icar...30..301G} (1977; hereafter GN) argued that resonant eddies on smaller scales damp the shear. There was later support for dissipation by resonant eddies in both analytic calculations \citep{1997ApJ...486..403G} and numerical simulations \citep{2011ApJ...731...67P, 2011ApJ...734..118P, 2012MNRAS.422.1975O}. Assuming Kolmogorov scalings for the turbulent eddies, GN argue for a quadratic reduction factor $(P_{\rm f}/2\pi \tau_\ell)^2$. For $P_{\rm f} \ll \tau_\ell$, this leads to a very large suppression of the viscosity. By contrast, even for resonant eddies, some numerical simulations \citep{2011ApJ...731...67P, 2011ApJ...734..118P} find a scaling closer to linear, perhaps due to the fact that the largest eddies do not follow inertial-range, Kolmogorov scalings. Other simulations \citep{2012MNRAS.422.1975O} find a quadratic scalings, with a negative value for the viscosity at high frequencies. As the RGB stars studied here may have turnover times $\ga 10^2\, \rm days$ while the orbital periods are of order days to weeks, there may be several orders of magnitude difference in the predictions given by un-reduced (here called ``standard" or ``std"), linear and quadratic scalings. 

The treatment of the equilibrium tide in the present paper was influenced by \citet{1992RSPTA.341...39P} and \citet{1995A&A...296..709V}, who studied circularization of binaries containing an RGB star. The latter presented analytic formulae for circularization which take into account the evolution of the RGB star to larger radius and luminosity. They applied this analytic circularization formula to 28 binaries, finding that all systems could be well explained with a non-reduced kinematic viscosity. Equivalent formulae are derived here for the orbital decay problem and for reduced viscosity. 

Binary orbital frequencies are below the frequencies of acoustic and fundamental waves, and hence only low-frequency gravity waves can be resonantly excited by the tide. During the SGB and RGB phase, the core is radiative and internal gravity waves are excited by the tide at the boundary between the radiative zone and the convective envelope \citep{1998ApJ...507..938G, 2016CeMDA.126..275B, 2013MNRAS.434.1079C, 2013MNRAS.432.2339I, 2017A&A...604A.112G, 2017ApJ...849L..11W, 2017MNRAS.467.2146K}. The waves then travel inward toward the center. Two damping mechanisms may prevent the formation of a standing wave. Radiative diffusion is dominant on the SGB and RGB, and can easily damp the dynamical tide from any orbiting companion, independent of its mass. Even if radiative diffusion were not present, companions larger than roughly 1 Jupiter mass will give rise to such large wave amplitudes near the center that the stratification is overturned and the wave breaks, depositing it's energy as heat and torquing the gas in the wave-breaking layer \citep{2010MNRAS.404.1849B,2011MNRAS.414.1365B}. In this paper it is assumed that the angular momentum is efficiently redistributed to the rest of the star. 
		
Numerous studies (e.g., \citealt{2008MNRAS.386..155S,2011ApJ...737...66K,2012ApJ...761..121M,2014ApJ...794....3V}) have considered the effect of post-main sequence stellar evolution and tidal evolution on planetary orbits, with the goal of predicting the properties of planetary systems around white dwarfs. For massive planets in close orbits, orbital decay can lead to engulfment of the planet. By contrast, smaller mass planets that do not suffer orbital decay must wait for the star's radius to expand out to their orbit. \citet{2012ApJ...761..121M} employed turbulent viscosity in the star's convective envelope as the tidal friction, and experimented with different prescriptions for reduction of the viscosity when the eddy turnover time is longer than the forcing period. However, since the semi-major axes explored were all large ($\ga 1$ AU), eddies in the convection zone turn over fast compared to the forcing period, and the full turbulent viscosity ends up being used. For those planetary systems that do not undergo a merger, high mass loss rates during the post-main sequence evolution will cause orbits to expand as the star's mass decreases.
	
The paper is organized as follows.
The prescriptions for the equilibrium and dynamical tide dissipation rates are discussed in Sections \ref{sec:eqtide} and \ref{Dynamical Tide Dissipation Rate}, respectively. Section \ref{sec:examples} shows examples of orbital evolution. Theory is compared with data from the APOGEE survey and the exoplanets in Section \ref{sec:acrit}. Conclusions are given in Section \ref{sec:conclusion}. Analytic approximations to the equilibrium tide dissipation rate and critical separation are given in Appendices \ref{sec:analytic_eq_tide} and \ref{sec:analytic_acrit}, respectively.

\section{Equilibrium Tide Dissipation Rate}
\label{sec:eqtide}

Consider a primary star of mass $M_1$ and radius $R_1$ in a circular orbit of separation $a$ with a secondary star of mass $M_2$. The tide raised by the secondary in the convective envelope of the primary creates a time-dependent fluid shear, which is damped by the turbulent viscosity of convective eddies. This process transfers energy and angular momentum from the orbit to the stellar convection zone. For simplicity, efficient angular momentum redistribution is assumed in the primary so that the rotation rate, $\Omega$, is uniform over the star. 
	
The orbital torque, $N$, is related to the energy dissipation rate in the rotating frame, $\dot{E}$, through the pattern speed $n-\Omega$ as $N=\dot{E}/(\Omega-n)$, where $n=\sqrt{ G(M_1+M_2)/a^3}$ is the orbital frequency. This relation is valid even as $\Omega \rightarrow n$, i.e. synchronous rotation. The torque $N \rightarrow 0$, since $\dot{E}\sim(\Omega-n)^2 \rightarrow 0$ as $\Omega \rightarrow n$. This torque changes the orbital angular momentum, $L=\mu n a^2$, where $\mu=M_1M_2/(M_1+M_2)$ is the reduced mass of the system. The semi-major axis then changes at a rate
\begin{equation}
	\label{dota_dotE}
	\dot{a} = - \frac{2\dot{E}}{\mu n (n - \Omega)a}.
	\end{equation}
Since $\dot{E}>0$, the orbit decays for a slowly rotating star ($\Omega < n$) and expands for a rapidly rotating star ($\Omega > n$). Equation~\ref{dota_dotE} agrees with the end result of the derivation in \cite{1981A&A....99..126H}, and is also valid even when the moment of inertia of the primary star is changing with time due to stellar evolution.

For massive secondaries, the spin of the primary may be tidally synchronized to the orbit, with subsequent tidal evolution occurring on the stellar evolution timescale of the primary (e.g., \citealt{2015A&A...574A..39D}).
The conserved total angular momentum of the system, assumed aligned, is $J=L + S_1 + S_2$, where $S_1=I_1\Omega$ is the spin angular momentum of the primary, and $S_2$ is that of the secondary. Here $I_1$ is the moment of inertia of the primary, and $\Omega$ is the primary's angular velocity. Since the moment of inertia of the secondary $I_2$ is much smaller than $I_1$, we ignore $S_2$. In that case, 
\be
L + S_1 & \simeq & \rm constant.
\label{eq:LplusS}
\ee
At each time $t$, $\Omega(t)$ is determined from the initial values $L(0)$ and $S_1(0)$ as $\Omega(t) = (L(0)+S_1(0)-L(t))/I_1(t)$. In this way a separate differential equation is not needed for $\Omega$. 

One technical point is that even a tiny amount of mass loss would  cause an artificial spinup of the primary according to Equation \ref{eq:LplusS}, since $L$ would decrease even at fixed $a$. To eliminate this issue, the two masses are fixed during the evolution. This is a good approximation for the stars near the base of the giant branch considered here, since little mass has been lost, and e.g., the expansion of the orbit would be tiny at this stage.

The viscous dissipation rate for incompressible flow is given by \citep{1959flme.book.....L}
\begin{equation}
	\label{eq:Edotvisc}
	\dot{E} = \frac{1}{2} \sum_{i=x,y,z} \sum_{k=x,y,z} \int \rho \nu \bigg( \frac{\partial v_i}{\partial x_k} + \frac{\partial v_k}{\partial x_i} \bigg)^2 d^3x
	\end{equation}
where the sums are over the three spatial directions, $v_i$ is the velocity of the tidal flow, $x_i$ are Cartesian coordinates, $\nu$ is the (isotropic) kinematic viscosity, and $\rho$ is the mass density. Isotropic turbulent viscosity is assumed for simplicity. Numerical simulations find modest deviations from isotropy for Boussinesq convection \citep{2009ApJ...705..285P,2011ApJ...734..118P}. The velocity of the tidal flow may be represented as a spherical harmonic expansion
\be
		\label{eq:vexpansion}
	\vect{v}   & = &  
        \sum_{\ell m} (-i \omega_m) \left[ \xi_{r, \ell m}(r)Y_{\ell m}(\theta,\phi)\vect{e}_r 
        \right. \nonumber \\ & + & \left. \xi_{h, lm}(r) r \vect{\nabla} Y_{\ell m}(\theta,\phi) \right]
        e^{-i \omega_m t}
	\ee
where $\xi_{r, \ell m}$ and $\xi_{h,\ell m}$ are the radial and horizontal component of the Lagrangian displacement vector, and $\omega_m = m(n-\Omega)$ is the forcing frequency in the rotating frame. 
Plugging Equation \ref{eq:vexpansion} into Equation \ref{eq:Edotvisc} and performing the angular integral gives
\begin{equation}
	\begin{split}
	\dot{E} & = \frac{1}{2} \sum_{\ell m}\omega_{m}^2 \int^{R_1}_{r_{\rm{bcz}}} dr r^2\rho \nu  \bigg[4\bigg( \frac{d\xi_{r, \ell m}(r)}{dr} \bigg)^2  \\
	 & + 2\ell(\ell+1)\bigg( \frac{d\xi_{h, \ell m}(r)}{dr} + \frac{\xi_{r, \ell m}(r)}{r} - \frac{\xi_{h, \ell m}(r)}{r} \bigg)^2  \\
	 & + 2\bigg(\ell(\ell+1)\frac{\xi_{h, \ell m}(r)}{r} - 2\frac{\xi_{r, \ell m}(r)}{r}\bigg)^2 \bigg]
	\end{split}
	\label{Edot_general}
	\end{equation}
where $r_{\rm bcz}$ is the radius at the base of surface convective zone.
Since the $m=0$ term has zero frequency it does not contribute. The $\pm m$ terms give equal contributions. 

The tidal potential in the primary due to a secondary orbiting at co-latitude $\pi/2$ and and orbit angle $(n-\Omega)t$ is	
\be
\begin{split}
U = & \sum_{\ell=2}^\infty \sum_{m=-\ell}^\ell U_{\ell m}(r) Y_{\ell m}(\theta,\phi) e^{-i\omega_{m}t} \\
= & -GM_2 \sum_{\ell=2}^\infty \sum_{m=-\ell}^\ell \frac{4\pi}{2\ell+1} \frac{r^{\ell}}{a^{\ell+1}}\\ 
& \times  Y_{\ell m}\left( \frac{\pi}{2},0 \right) Y_{\ell m}(\theta,\phi) e^{-i\omega_{m}t},
\end{split}
\ee
which is smaller than the potential $GM_1/R_1$ at the surface of the primary by a small factor $\epsilon = (M_2/M_1)(R_1/a)^3$ for $\ell=2$.
In the equilibrium tide approximation (e.g., \citealt{1977Icar...30..301G}), the radial and horizontal displacements are
\begin{equation*}
	\xi_{r, \ell m}(r) = -\frac{U_{\ell m}(r)}{g}
\end{equation*}
and
\begin{equation*}
	\xi_{h, \ell m}(r) = -\frac{1}{\ell(\ell+1)} \frac{ U_{\ell m}(r) }{ g} \bigg( 2\ell - \frac{d\ln g}{d\ln r} \bigg)
\end{equation*}
Given the run of $\rho$, $\nu$, $g$, $\xi_{r,\ell m}$ and $\xi_{h,\ell m}$ versus $r$ for a stellar model, $\dot{E}$ is computed by numerical integration of Equation \ref{Edot_general} for a given stellar model. Analytic approximations for this integral are discussed in the Appendix. 

When the turnover time $\tau_{\rm ed}$ of the large eddies becomes longer than the forcing period $P_f=P_{\rm orb}/2$ (for $m=2$), it is expected than turbulent viscosity is reduced, since the large eddies cannot transport momentum efficiently.  Three models of turbulent viscosity will be investigated: 
un-reduced (``standard") viscosity
\be
\nu_{\rm std} & = & \frac{1}{3} v_{\rm ed} \alpha H, 
\label{eq:nustd}
\ee
Zahn's formula with a linear reduction
\be
\nu_{\rm Z} & = & \nu_{\rm std} \times {\rm min} \left( 1 ,  \frac{P_f}{2\tau_{\rm ed}} \right),
\label{eq:nuZ}
\ee
and Goldreich and Nicholson's formula with a quadratic reduction
\be
\nu_{\rm GN} & = & \nu_{\rm std} \times {\rm min} \left( 1 ,  \left( \frac{P_f}{2\pi \tau_{\rm ed}} \right)^2 \right),
\label{eq:nuGN}
\ee
The numerical factors in each expression are somewhat arbitrary \citep{2009ApJ...705..285P}.
Due to the large eddy velocity near the surface, the eddy turnover time will generally be shorter than forcing periods of interest there. Reduced viscosity is important deep in the convection zone where eddy velocities are small, due to increasing density. 

If the viscosity scales with orbital period $P_{\rm orb}=2\pi/n$ as $\nu \propto P_{\rm orb}^\alpha$, then for $\Omega \ll n$, $\dot{E} \propto M_2^2 a^{-9+3\alpha/2}=M_2^2 a^{-(\beta+2)}$ and 
	$\dot{a} \propto M_2 a^{-7+3\alpha/2}=M_2 a^{-\beta}$ with $\beta=7-3\alpha/2$. The relevant values of $\beta$ are then 7, 5.5 and 4 for un-reduced, linear and quadratic scalings. The exponent $\beta$ is a crucial parameter that directly determines the relative number of systems at different orbital separation.

Appendix \ref{sec:analytic_eq_tide} discusses analytic approximations to the dissipation rate for the different turbulent viscosity assumptions. The convective envelope is treated as an $n=3/2$ polytrope, with the entropy increasing as the star ascends the RGB. These approximations will be used in Section \ref{sec:acrit} to understand the critical orbital separation out to which orbital decay is expected to have caused systems to merge.

\subsection{A Numerical Example}

\begin{figure}
\centering
	\includegraphics[width=0.5\textwidth]{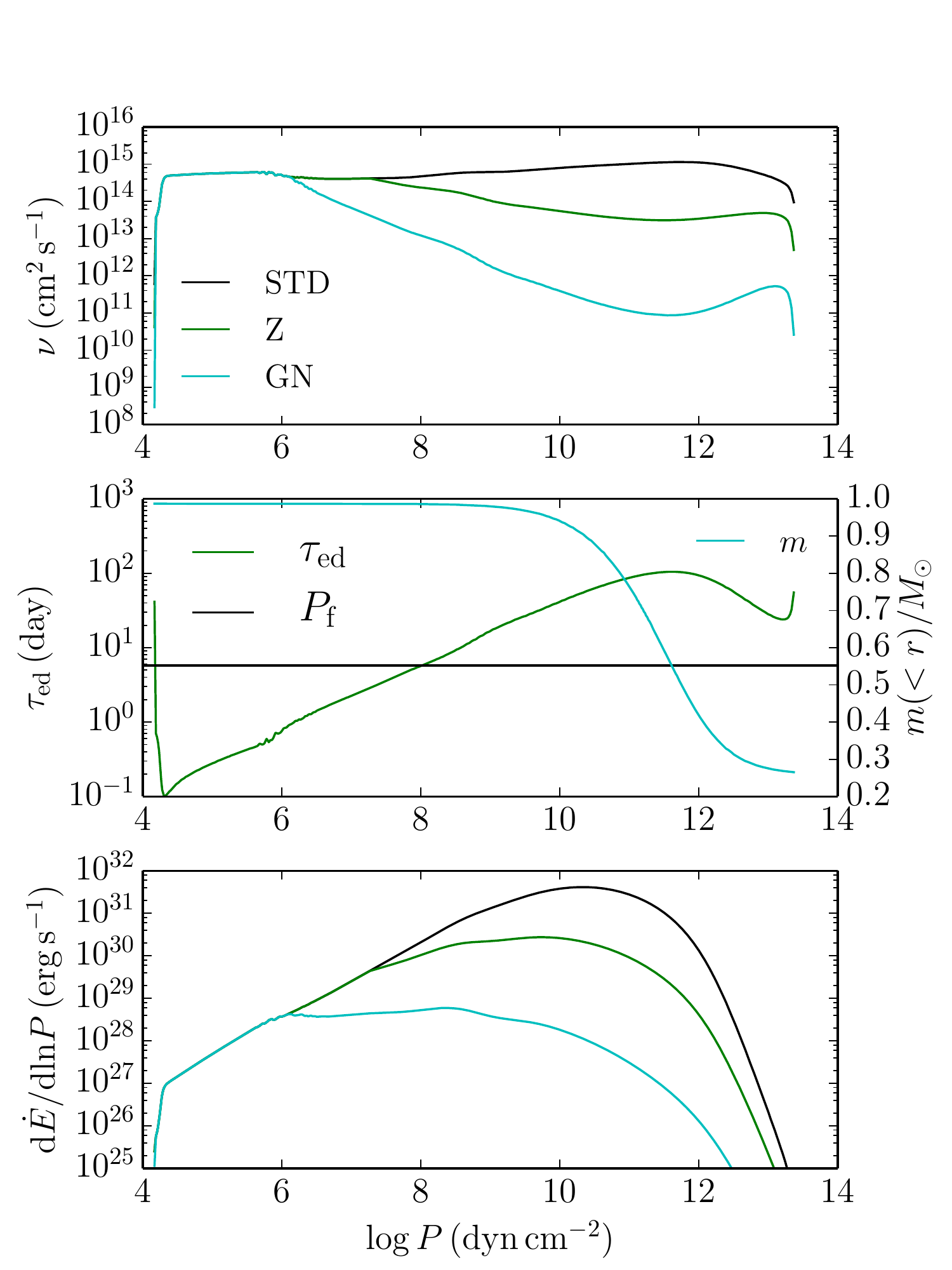}
	\caption{Depth dependence of quantities needed for the turbulent viscosity dissipation rate in the convective envelope. The top panel shows the run of the three prescriptions for turbulent viscosity as a function of pressure in the convective envelope. The middle panel shows the eddy turnover time (green solid line), forcing period ($P_{\rm orb}/2=5.7\, \rm days$, the black horizontal line) and interior mass $m(r)$ in terms of ${\rm log}\,P$. The bottom plot gives the integrand of the energy dissipation integral. The top and bottom panels show the un-reduced, Zahn and GN prescriptions as black, green and cyan lines, respectively. 
	The base of the convective zone is on the right of the figure, where $\log\,P = 13.5$. The parameters used are a $M_1=1\, M_{\odot}$ RGB star with radius $R_1 = 10\, R_{\odot}$, companion mass $M_2=0.01\, M_{\odot}$, and separation $a = 0.1$ AU (orbital period $P_{\rm orb}=11.4$ days).}
	\label{fig:viscosity_energy_dissipation_logP}
\end{figure}

For close-in orbits and deep stellar convection zones, the viscosity is expected to be significantly reduced. This section shows a numerical example to illustrate the reduction factor. Figure \ref{fig:viscosity_energy_dissipation_logP} shows the depth dependence of turbulent viscosity (top panel), eddy turnover time and interior mass (middle panel), and $d\dot{E}/d\ln P$ from Equation \ref{Edot_general} (bottom panel); here $P$ is pressure. Representative values have been used, with a $M_1=1M_{\odot}$ RGB star with radius $R_1=10R_{\odot}$, companion mass $M_2=0.01\, M_{\odot}$, orbital separation $a=0.1\, \rm AU$ and orbital period $P_{\rm orb}=11.4$ days.

The middle panel of Figure \ref{fig:viscosity_energy_dissipation_logP} shows that this RGB star has eddy turnover times $\tau_{\rm ed} =  \alpha_{\rm mlt} H/v_{\rm ed} \ga 100\, \rm days$ near the peak of the integrand. Hence tidal forcing for any orbit shorter than $P_{\rm orb}\simeq 200\, \rm days$ is expected to have reduced viscosity. The peak in eddy turnover time is due to the scale height becoming small both near the surface, due to lower temperature, and toward the center, due to higher gravity. 

Near the peak of the integrand the eddies are turning over 10 times slower than the tidal forcing period. The top panel shows that this leads to a reduction factor of $10^{-1}$ and $10^{-4}$ relative to un-reduced viscosity, for the Zahn and GN viscosities, respectively. The bottom panel shows that $d\dot{E}/d\ln P$ has a strong peak for un-reduced viscosity, due to decreasing mass near the surface and decreasing tidal shear toward the interior. The integrands for Zahn and GN viscosities are reduced in size and with flattened peaks closer to the surface. The Zahn case still has a peak in the integrand well below the depth where $P_f = \tau_{\rm ed}$, however the GN case has a flat integrand over 3-4 decades in pressure, ending at $P_f = \tau_{\rm ed}$.

\section{ Dynamical Tide Dissipation Rate}
\label{Dynamical Tide Dissipation Rate}

The dynamical tide involves the excitation of internal gravity waves near the radiative-convective boundary (RCB). If damping is weak, waves will reflect in the core and form standing waves. When the wave can be damped in less than one group velocity travel time the result is a traveling wave. In the traveling wave regime, the dissipation rate is given by the inward-going wave luminosity, $L_{\rm dyn}$.

During the SGB and RGB phases the core is radiative and the envelope convective. The dynamical tide is excited at the RCB, with the waves propagating inward toward the center (e.g., \citealt{1998ApJ...507..938G}). There are two damping mechanisms that may cause the wave to damp at or before it reaches the center. First, radiative diffusion damping becomes progressively more important as the star evolves off the MS, due to the large number of internal gravity wave nodes in the core \citep{1977AcA....27...95D}. This mechanism depends only on the orbital period, and is independent of the companion mass. Second, waves may ``break" nonlinearly at the center, and not reflect back \citep{2010MNRAS.404.1849B,2011MNRAS.414.1365B}. Here wave breaking may mean overturning the local stratification, or strong wave-wave interactions which transfer energy from the tidally-excited fluid motion to daughter waves \citep{2012ApJ...751..136W,2010MNRAS.404.1849B,2011MNRAS.414.1365B, 2016ApJ...816...18E}. Nonlinear wave breaking depends on the companion mass, as well as the orbital period.


The traveling wave luminosity is computed from stellar models by solving the linearized radial momentum and continuity equations and applying the appropriate boundary conditions for an inward-going traveling wave in the radiative zone. Define the radial Lagrangian displacement $\xi_r$, the potential $\psi = \delta p/\rho + U$, the Eulerian pressure perturbation $\delta p$, and the tidal potential $U$.  The radial momentum equation and continuity equations, in the Cowling approximation, are then (e.g., \citealt{1989nos..book.....U})
\be
\frac{d\psi}{dr} & = & \frac{N^2}{g} \left( \psi - U \right) - (N^2-\sigma^2) \xi_r
\ee
and
\be
\frac{d\xi_r}{dr} & = & \xi_r \left( \frac{g}{c^2} - \frac{2}{r} \right)
+ \psi \left( \frac{k_h^2}{\sigma^2} - \frac{1}{c^2} \right) + \frac{U}{c^2},
\ee
where the horizontal wavenumber is $k_h^2=\ell(\ell+1)/r^2 $. At the surface, the evanescent wave boundary condition is \citep{1989nos..book.....U} $\delta p/\rho = \psi - U = g\xi_r$. The inward-going traveling wave boundary condition is
\be
\frac{d\left( \psi - \psi_0 \right)}{dr} & = & i k_r \left( \psi - \psi_0 \right),
\ee
where $\psi_0$ is an approximate long-wavelength, particular solution and $k_r=k_hN/\omega$ is the radial wavenumber. The value of $\psi_0$ can be computed from the equilibrium tide as
\citep{2010ApJ...714....1A,2012ApJ...751..136W, 2017ApJ...849L..11W}
\be
\psi_0 & = & \frac{\omega^2}{\ell(\ell+1)} \frac{d(r^2\xi_{\rm r,eq})}{dr}.
\ee
The dynamical tide pieces of $\xi_r$ and $\psi$ are denoted $\xi_{\rm r,dyn} = \xi_r - \xi_{\rm r,eq}$ and $\psi_{\rm dyn} = \psi - \psi_0$. In the gravity wave propagation zone, the traveling wave luminosity is given by 
\begin{equation}
L_{\rm dyn} = r^2 \int d\Omega \rho \psi_{\rm dyn} \dot{\xi}_{\rm r,dyn},
\label{L_dyn_numerical}
\end{equation}
which involves an integral over two spherical harmonics. The purpose of subtracting the long-wavelength response is to decrease the size of the oscillatory part of $L_{\rm dyn}(r)$, making it easier to isolate the value in the propagation zone far away from the RCB.

The decrease of gravity wave energy in the core due to radiative diffusion can be parametrized as 
\be
\alpha & \equiv & \frac{\dot{E}_{\rm diff}}{L_{\rm dyn}} \simeq 2 \int_0^{r_{\rm rcb}} dr \frac{k_h^3 N}{\omega^4} \nabla_{\rm ad}(\nabla_{\rm ad}-\nabla)\frac{4\sigma T^4 g^2}{3\kappa P^2}
\nonumber \\ & \equiv & \left( \frac{P_{\rm orb}}{P_{\rm orb, diff}} \right)^4.
\label{eq:alpha}
\ee
Here $\dot{E}_{\rm diff}$ is the dissipation rate including both the inward and outward going waves, giving the factor of 2 in Equation \ref{eq:alpha}.
The adiabatic and stellar temperature gradients are denoted $\nabla_{\rm ad}$ and $\nabla$, respectively, and $\kappa$ is the opacity. This formula was derived for the dynamical tide component using the low frequency, quasi-adiabatic approximation discussed in \citet{1989nos..book.....U}, and is valid even in the degenerate core and non-degenerate burning shell, where most of the contribution arises.

\begin{figure}
\centering
\includegraphics[width=0.5\textwidth]{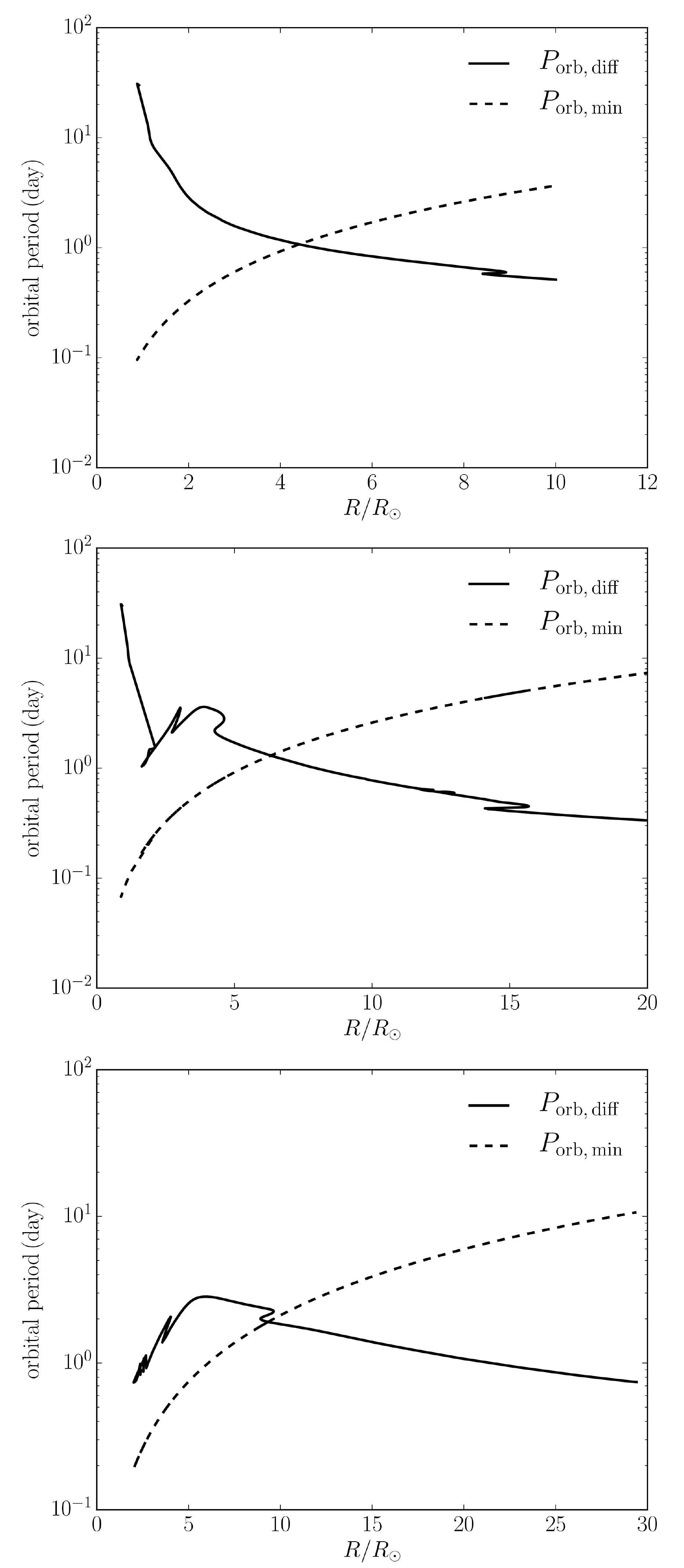}
\caption{ Critical orbital period during the evolution for stars of mass $M_1=1\,\,(\rm top\,\,panel),2\,\,(\rm middle\,\,panel),3\, M_\odot$ (bottom panel). The solid line shows the critical orbital period, $P_{\rm orb,\,diff}$ outside of which the radiative diffusion damping timescale is shorter than twice the group travel time across the core, and the traveling wave regime obtains. The dashed lines show the minimum possible orbital period $P_{\rm orb,\,min}$ for a star of that mass and radius, where the companion is orbiting at the surface of the star. At a given stellar radius, the standing wave limit can only occur if the dashed line is below the solid line.}
\label{fig:Porbcritdiff}
\end{figure}

Outside $P_{\rm orb, diff}$, defined in Equation \ref{eq:alpha},  radiative diffusion damping is strong, and the traveling wave limit occurs, and vice versa for $P_{\rm orb} < P_{\rm orb, diff}$. Figure \ref{fig:Porbcritdiff} shows $P_{\rm orb,\,diff}$ versus stellar radius, $R$, which acts as a proxy for age. In all three cases, radiative diffusion will lead to the traveling wave limit for orbital periods $P_{\rm orb} \ga 1\, \rm day$ over part of the SGB and all the RGB. This enhanced radiative diffusion is caused by the short vertical wavelength in the core.

Radiative diffusion may also have an important effect on the driving of waves at the RCB. For very thin surface convection zones in MS stars of mass $M_1 \ga 1.3\, M_\odot$, the thermal time can become shorter than the forcing period, and radiative diffusion will rapidly damp out temperature differences induced by the wave, effectively eliminating the buoyancy force. The wave luminosity is expected to be suppressed when $\omega \la \omega_{\rm diff}$, where the 
thermal diffusion frequency at the RCB is approximated as
\be
\omega_{\rm diff} & \simeq & \frac{F}{\rho g \lambda^2}.
\label{omega_diff}
\ee
Here $F$ is the flux and $\lambda \equiv|\ell(\ell+1)(dN^2/dr) / (\omega^2r^2)|^{-1/3} \simeq H (\omega/c_sk_h)^{2/3}$ is the Airy wavelength of the gravity wave \citep{1998ApJ...507..938G}. 
In this paper it is assumed that rapid thermal diffusion will greatly reduce the wave luminosity, so that it is set to zero when 
the thermal time $t_{\rm th}=PC_pT/gF$ is shorter than the forcing period $P_{\rm f}$. As the convective envelope rapidly deepens on the SGB, where $\omega_{\rm diff}$ decreases fast and $t_{\rm th}$ increases fast, $\omega = \omega_{\rm diff}$ and $t_{\rm th}=P_{\rm f}$ occur at almost the same stage during the evolution. Therefore, turning on the dynamical tides at $t_{\rm th}>P_{\rm f}$ is a good approximation. This effectively sets the inward-going wave luminosity to zero on the MS for the $M_1=2$ and $3\, M_\odot$ cases. As the star leaves the MS, the convection zone will deepen rapidly and hence our assumption allows the dynamical tide dissipation to turn on suddenly at the end of the MS. This effect is clearly evident in the results for the $M_1=2$ and $3\, M_\odot$ stars.

In the calculations of orbital decay in later sections, it is convenient to have tabulated formulas for the wave luminosity that can be rapidly evaluated, as opposed to solving the above boundary value problem. If $\lambda \simeq \left( \omega / ck_h \right)^{2/3} H$ is much smaller than the local scale height $H$, which is valid when the forcing frequency is much smaller than the Lamb frequency at the RCB,  the Airy approximation is good and the wave luminosity can be written in the form \citep{1998ApJ...507..938G}
\begin{equation}
\begin{split}
L_{\rm dyn} = & \bigg(\frac{3^{2/3}\Gamma^2(1/3)}{2\pi}\bigg)[\ell(\ell+1)]^{-4/3}\omega^{11/3}\\
& \times \bigg(\rho r^5 \bigg|\frac{dN^2}{d{\rm ln}r}\bigg|^{-1/3}\frac{\zeta^2\xi^2_{r,{\rm eq}}}{r^2}\bigg),
\end{split}
\label{L_dyn_analytical}
\end{equation}
where the dimensionless parameter $\zeta$ is defined by
\be
\frac{d\xi_{\rm r,dyn}}{dr} & \equiv & \zeta \frac{\xi_{\rm r,eq}}{r},
\ee
and $\zeta$ is set by matching the solution in the convection zone to that in the radiative zone. All the quantities in Equation \ref{L_dyn_analytical} are evaluated at the RCB. MESA models are used to compute $\zeta$ as a function of age for each stellar model. Our numerical calculations find that $\zeta$ grows strongly during the RGB.
Equating the analytical formula in Equation \ref{L_dyn_analytical} to the numerical results generated from Equation \ref{L_dyn_numerical} for each stellar model gives the parameter $\zeta$. When both dynamical and equilibrium tides are included, the orbital decay rate becomes $\dot{a} = - 2(\dot{E}_{\rm eq} + |L_{\rm dyn}|)/ \mu n (n - \Omega)a$.

When applying Equation \ref{L_dyn_analytical} in the calculations of orbital decay, the use of the Airy approximation on the radiative side of the RCB requires that the wavelength is always much shorter than a scale height. For closer orbits, the larger forcing frequency implies larger wavelengths, and the luminosity can be larger than implied by Equation \ref{L_dyn_analytical}, and this approach may underestimate the orbital decay rate.


Given $L_{\rm dyn}$, the nonlinearity of the wave must be checked by evaluating
\begin{equation}
k_r \xi_{r,{\rm dyn}}=\sqrt{\frac{[\ell(\ell+1)]^{3/2} N L_{\rm dyn}}{4\pi\rho r^5\omega^4}}
\end{equation}
in the radiative zone. If $\alpha > 1$, or if $\alpha < 1$ but  $k_r \xi_{r,{\rm dyn}}> 1$, the dissipation rate is given by the full $|L_{\rm dyn}|$. On the other hand, if $\alpha < 1$ and $k_r \xi_{r,{\rm dyn}}<1$, the dissipation rate is set to zero as the wave reflects back and forms a standing wave, with much smaller dissipation rate.

Lastly, we have ignored the Coriolis force in computing the dynamical or quasi-static tidal response. This is not a good approximation when the rotation is nearly synchronized, as occurs for sufficiently massive companions. As the forcing frequency $\sim 2(n-\Omega)$ becomes small, inertial waves may be excited, possibly giving much larger dissipation rates than included here (e.g. \citealt{2007ApJ...661.1180O}). For SGB and RGB stars, the large expansion after the MS implies these stars will be slowly rotating before tidal torques become important. In this case, the bottleneck is first spinning the stars up to synchronous rotation. At that point, the orbital decay rate becomes small and further evolution is on the nuclear timescale of the primary. Increasing the dissipation rate at fixed $n-\Omega$ will tend to make $n-\Omega$ smaller, until the synchronization rate becomes comparable to the nuclear timescale of the primary. Hence we believe including the Coriolis force and inertial waves will not significantly change our results for orbital decay.

\section{ Examples of Orbital Decay}
\label{sec:examples}

This section presents calculations of orbital decay for a range of primary mass, secondary mass and initial separation. Each integration of the equation $\dot{a} = - 2(\dot{E}_{\rm eq} + |L_{\rm dyn}|)/ \mu n (n - \Omega)a$ includes the dynamical tide, as well as a prescription for the turbulent viscosity used to damp the equilibrium tide. Calculations using Zahn and Goldreich-Nicholson viscosity are compared to assess if they result in potentially detectable differences in the critical separation for rapid orbital decay. The range of substellar companion masses is chosen to span the range of synchronized and non-synchronized cases. The range of primary masses and post-MS evolutionary state represent the bulk of binaries with SGB and RGB primaries and a substellar secondary.

Modules for Experiments in Stellar Astrophysics (MESA, {version 8845},  \citealt{2011ApJS..192....3P,2013ApJS..208....4P,2015ApJS..220...15P}) is used to provide the stellar structure for three stars of mass  $M_1=1, 2$ and $3\, M_{\odot}$ during the MS, and ending on the RGB. The initial metallicity is $Z=0.02$. The type 2 opacity table is used. The nuclear burning network used is ``o18\_and\_ne22.net". The mixing length factor is 2. The Schwarzschild criterion for the definition of the convective zone is applied.

In the following three sections, results are presented for the three different primary masses.

\subsection{$1\, M_\odot$ Model}

Sun-like stars have a radiative core and a relatively thick convective envelope on the MS, allowing the ingoing-wave dynamical tide to operate. The convection zone deepens significantly on the RGB in both mass and radius, giving rise to equilibrium tide dissipation rates many orders of magnitude larger than on the MS. At fixed semi-major axis, the dynamical tide increases strongly during the SGB, and is relatively constant on the RGB.
	
\begin{figure}
\centering
\includegraphics[width=0.5\textwidth]{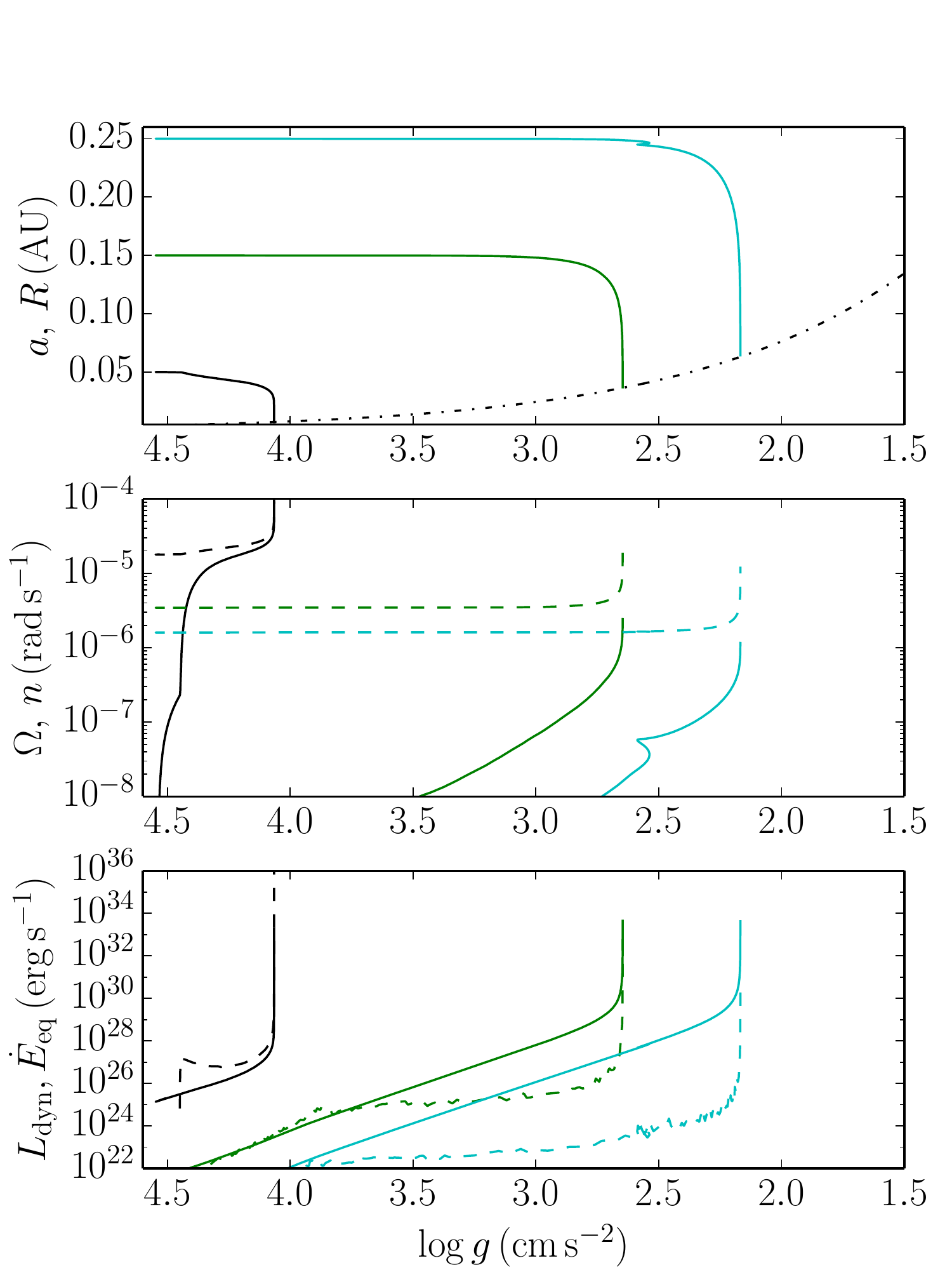}
\caption{Orbital decay for a $M_1=1\, M_\odot$ primary, with companion mass $M_2= 0.01\, M_{\odot}$ and initial separations $a=$ 0.05 (black), 0.15 (green) and 0.25 (cyan) AU. Zahn's turbulent viscosity prescription ($\nu_{\rm Z}$) has been used for the equilibrium tide. Here $\log g$ of the primary star shows the evolution of the star, from left to right.  (Top panel) Semi-major axis vs $\log g$. The black dotted-dash line shows the stellar radius, $R_1$. (Middle panel) Primary rotation rate ($\Omega$, solid lines) and binary orbital frequency ($n$, dashed lines). (Bottom panel) Dynamical tide ($|L_{\rm dyn}|$, dashed lines) and equilibrium tide ($\dot{E}_{\rm eq}$, solid lines) dissipation rates. }
\label{1M_10MJ_ZN}
\end{figure}

The top panel of Figure~\ref{1M_10MJ_ZN} shows the evolution of $a$ for a $M_1=1M_{\odot}$ primary and a $M_2=0.01M_{\odot}$ secondary, including dynamical tides and equilibrium tides with Zahn's prescription ($\nu_{\rm Z}$) for viscosity. The surface gravity, $\log\,g\equiv \log_{10}(g/(\rm cm\,s^{-2}))$ is a proxy for the evolutionary state of the primary star from the MS phase ($\log\,g =4.5 $) to the RGB ($\log\,g \ll 4.5$). When the orbital decay rate is small, $a$ is constant. Since the orbital decay rate increases so strongly after the MS, the orbit will decay rapidly compared to the stellar evolution timescale, and the line will become nearly vertical. The system merges at the end of the MS for $a_{\rm init}=0.05$ AU, 
and during the RGB phase for $a_{\rm init}=0.15$ and $0.25$ AU.
From the middle panel, none of the examples synchronize for this relatively low companion mass (starting from $\Omega=0$). The bottom panel shows the dynamical tide wave luminosity $L_{\rm dyn}$ and the equilibrium tide energy dissipation rate $\dot{E}_{\rm eq}$ during the evolution. For the case of  $a_{\rm init}=0.05$ AU, the dynamical tide dominates.
For the case of $a_{\rm init}=0.15$, the dynamical and equilibrium tides are comparable before the RGB phase, then the equilibrium tide increases much faster than the dynamical tide. In the case of $a_{\rm init}=0.25\, $AU, the orbit shrinks mainly due to the equilibrium tide. 

There are two trends that favor the equilibrium tide over the dynamical tide for wider orbits. First, they have different dependence on orbital period, with $\dot{E}_{\rm eq} \propto P_{\rm orb}^{-5.5}$ and $L_{\rm dyn} \propto P_{\rm orb}^{-7.67}$,
so the former decreases outward more slowly. Second, the dynamical tide luminosity has an initial increase by several orders of magnitude during the SGB, but then becomes relatively constant during the RGB (at fixed semi-major axis). This is in contrast to the equilibrium tide, which shows a continuous increase up the RGB. Hence for  decay of  wider orbits, which occurs for a more evolved primary star, the equilibrium tidal friction is more important.

\begin{figure}
\centering
\includegraphics[width=0.5\textwidth]{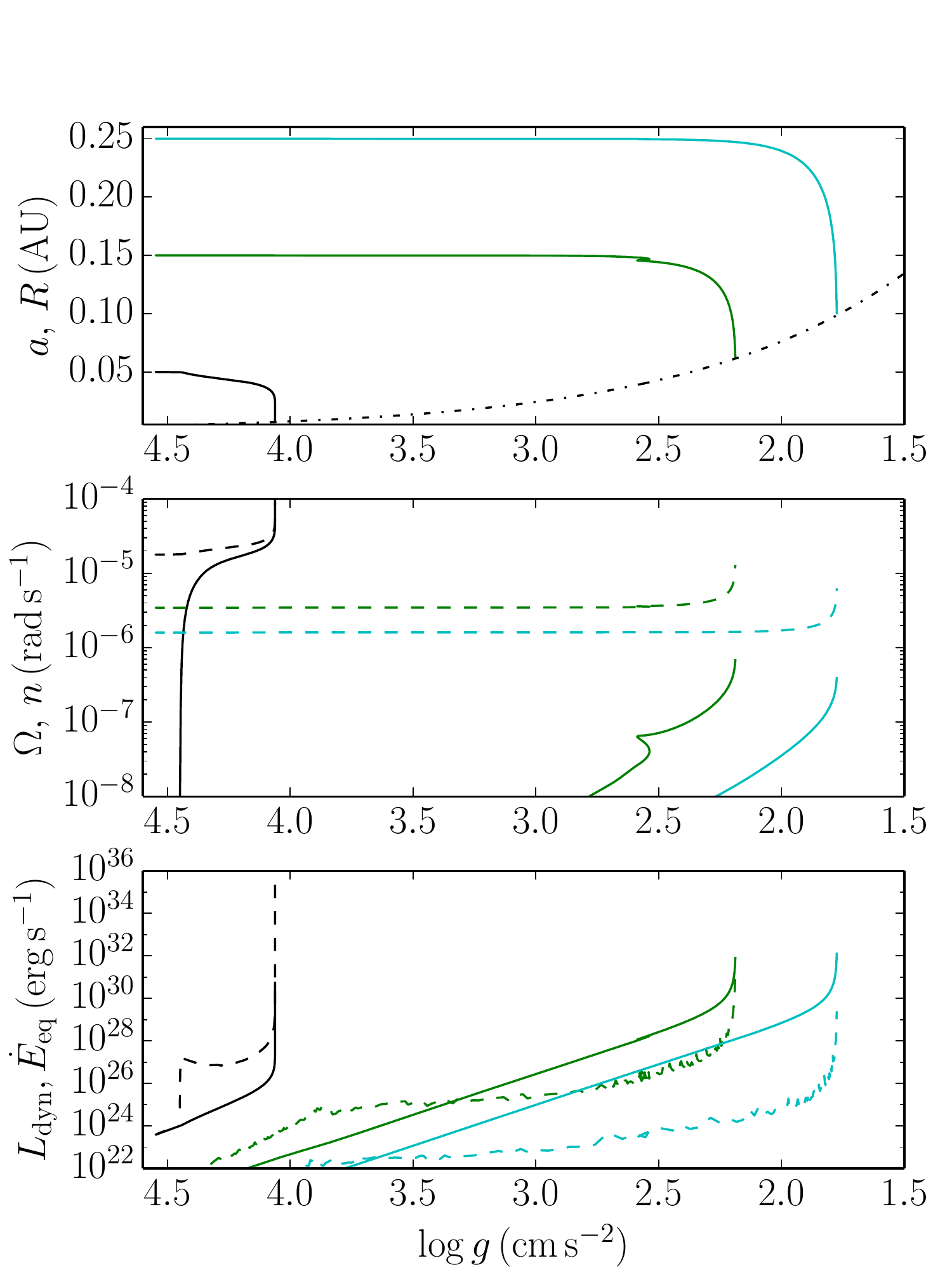}
\caption{ Same as Fig.~\ref{1M_10MJ_ZN} but using GN's viscosity ($\nu_{\rm GN}$) for the equilibrium tide dissipation.
}
\label{1M_10MJ_GN}
\end{figure}

Figure \ref{1M_10MJ_GN} is the same as Figure \ref{1M_10MJ_ZN} but uses GN's viscosity prescription instead of Zahn's. 
The equilibrium tide dissipation rate is then much smaller since $\nu_{\rm GN} \ll \nu_{\rm Z}$.
The $a=0.05\, \rm AU$ orbit decays at a similar time as in Figure \ref{1M_10MJ_ZN} since the dynamical tide dominates.
It also dominates early in the evolution for the $a=0.15\, \rm AU$ case, although there is little orbital decay during that time.


\begin{figure}
\centering
\includegraphics[width=0.5\textwidth]{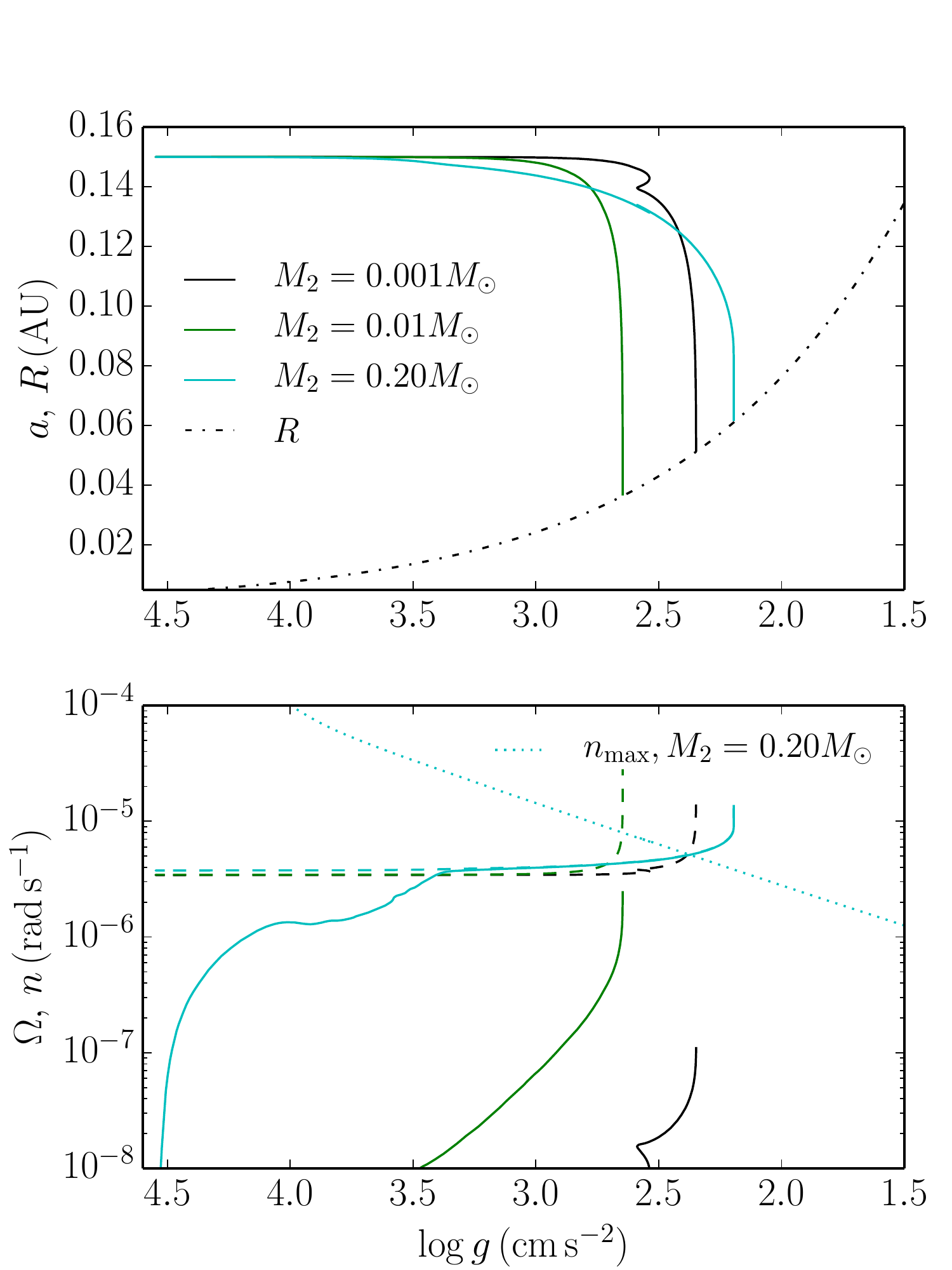}
\caption{The effect of synchronous rotation on orbital decay for $M_1=1\, M_\odot$, $M_2=0.001$ (black lines), $0.01$(green lines) and $0.2\, M_\odot$ (cyan lines). Zahn's viscosity is used.
(top panel) Semi-major axis ($a$) versus evolutionary state of the primary ($\log g$), and stellar radius. (bottom panel) Primary rotation rate ($\Omega$, solid lines), orbital frequency ($n$, dashed lines) and orbital frequency at which the Darwin instability begins ($n_{\rm max}$, dotted line), evaluated for secondary mass $M_2=0.2\, M_\odot$.}
\label{1M_015_big_companion_ZN.pdf}
\end{figure}

Next, Figure \ref{1M_015_big_companion_ZN.pdf} compares the orbital decay for a $M_1=1\, M_\odot$ primary with three different companion masses with Zahn's reduced viscosity. For the low mass companion, $M_2=10^{-3}\, M_{\odot} \simeq 1\, M_{\rm Jup}$, $\dot{a} \propto M_2$ leads to a small orbital decay rate, and the spin is far from synchronous. By comparison, the $M_2=0.01M_{\odot}$ case is also not synchronized, but the orbital decay occurs faster due to the larger mass. The higher mass, $M_2=0.2\, M_\odot$ case would have had even stronger orbital decay if synchronization did not occur. However, this system synchronizes on the SGB at $\log\,g=3.5$ by the equilibrium tide, after which point the orbit evolves on the much slower stellar evolution timescale, so that this case actually lives {\it longer} than the two lower mass cases. The lower panel of Figure \ref{1M_015_big_companion_ZN.pdf} shows that destruction occurs due to the Darwin instability, at separation $a_{\rm D} \simeq \left( 3I_1/\mu \right)^{1/2} \simeq \left(3I_1/M_2 \right)^{1/2}$ and orbital frequency $ n_{\rm max} \simeq \sqrt{G(M_1+M_2)/a_{\rm D}^3}\simeq \sqrt{GM_1/a_{\rm D}^3}$. \footnote{$a_{\rm D}$ can be derived by solving for the separation where $\partial J/\partial n =0$ assuming $\Omega=n$.} Beyond this point, the spin has larger angular momentum than the orbit, and the orbital frequency will tend to increase rapidly, leaving the spin frequency behind. Hence the critical semi-major axis out to which orbits are destroyed is not a monotonic function of $M_2$; for small $M_2$ it increases (see Appendix \ref{sec:analytic_acrit}), while for large $M_2$ it decreases. 

\begin{figure}
\centering
\includegraphics[width=0.5\textwidth]{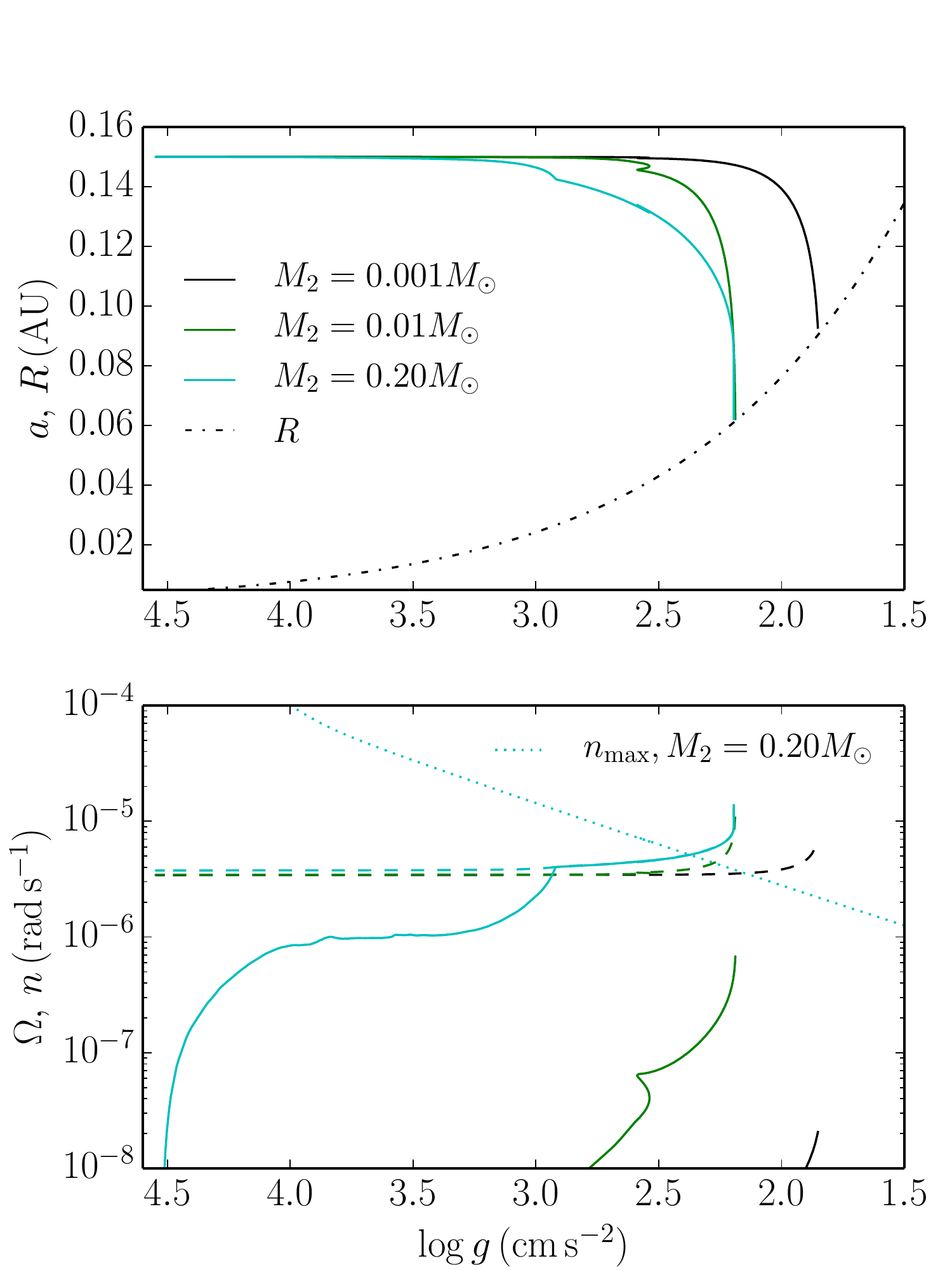}
\caption{Same as Figure \ref{1M_015_big_companion_ZN.pdf} but with GN's reduced viscosity.
}
\label{1M_015_big_companion_GN.pdf}
\end{figure}

Figure \ref{1M_015_big_companion_GN.pdf} is the same as Figure \ref{1M_015_big_companion_ZN.pdf} but uses GN's reduced viscosity prescription. For the $M_2=0.001M_{\odot}$ and $M_2=0.01M_{\odot}$ cases, the systems merge later than for $\nu_{\rm Z}$ since $\nu_{\rm GN} \ll \nu_{\rm Z}$. As the primary is not synchronized in both cases, the system merges earlier for the higher mass companion. For the $M_2=0.2M_{\odot}$ high mass companion, the orbit is synchronized later than in Figure \ref{1M_015_big_companion_ZN.pdf}, but the merger occurs at a similar time
due to the Darwin instability.



\subsection{2 $M_{\odot}$ Model}

For $M_1=2\, M_{\odot}$, the star has a convective core on the MS which inhibits nonlinear wave breaking at the center.\footnote{ An outward traveling wave flux excited at the radiative-convective boundary of the central convection zone is ignored in this paper.} More importantly, the surface convection zone is so thin that radiative diffusion damping suppresses the driving of the wave (see Equation \ref{omega_diff} and the surrounding discussion). On the SGB, the core becomes radiative and the surface convection zone deepens, at which point both efficient excitation and nonlinear wave breaking can occur.  

\begin{figure}
	\centering
	\includegraphics[width=0.5\textwidth]{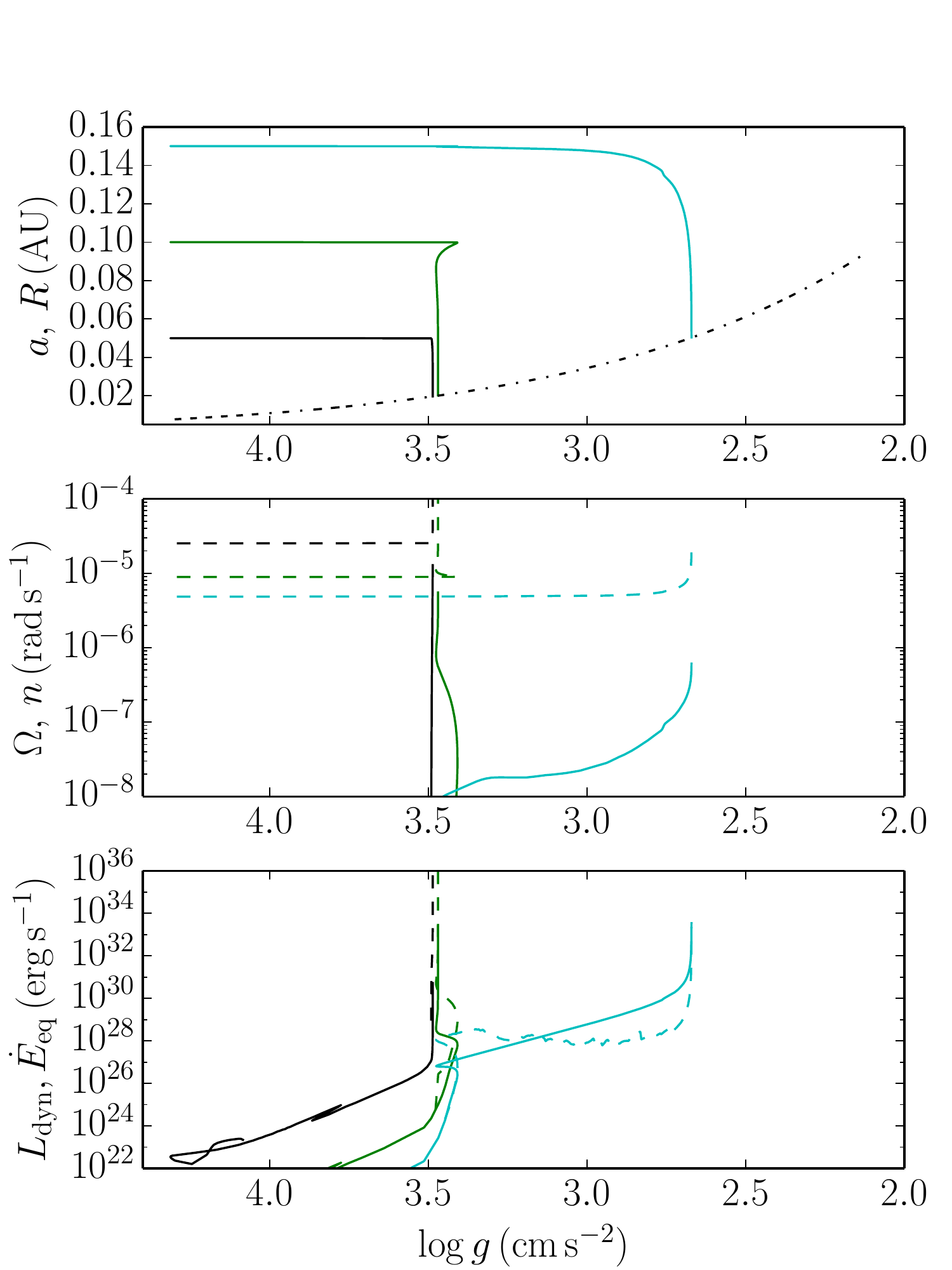}
	\caption{Same as Figure \ref{1M_10MJ_ZN} but with $M_1=2\, M_\odot$, $M_2=0.01\, M_\odot$ and $a_{\rm init}=0.05$ (black lines), 0.10 (green lines) and 0.15 (cyan lines) AU.  Zahn's turbulent viscosity prescription ($\nu_{\rm Z}$) has been used for the equilibrium tide. }
	\label{2M_10MJ_ZN.pdf}
\end{figure}

Figure \ref{2M_10MJ_ZN.pdf} shows the evolution for $M_1=2\, M_{\odot}$ and $M_2=0.01M_{\odot}$ starting from $a_{\rm init}=$ 0.05, 0.10 and 0.15 AU. Zahn's viscosity is used, and the dynamical tide is included. The bottom panel shows that $L_{\rm dyn}=0$ until $\log g=3.5$, where the surface convection zone deepens and $t_{\rm th}>P_{\rm f}$. Further $\dot{E}_{\rm eq}$ is much smaller than the $M_1=1\, M_\odot$ case on the MS, due to the smaller convection zone. The end result is that tidal friction in this model is much weaker than the $M_1=1\, M_\odot$ model on the MS, but comparable on the SGB and RGB. The middle panel shows that none of the cases synchronize for this companion mass. 
In the bottom panel, $L_{\rm dyn}$ dominates by orders of magnitude in the $a=0.05\, \rm AU$ and $a=0.1$ AU cases once it turns on. This causes the immediate decay of the orbit, on a shorter timescale than the radius expansion of the primary, seen as vertical lines in the top panel. For the $0.15\, \rm AU$ case, dynamical tides are important only briefly after they turn on, and then the equilibrium tide dominates until the merger.

\begin{figure}
	\centering
	\includegraphics[width=0.5\textwidth]{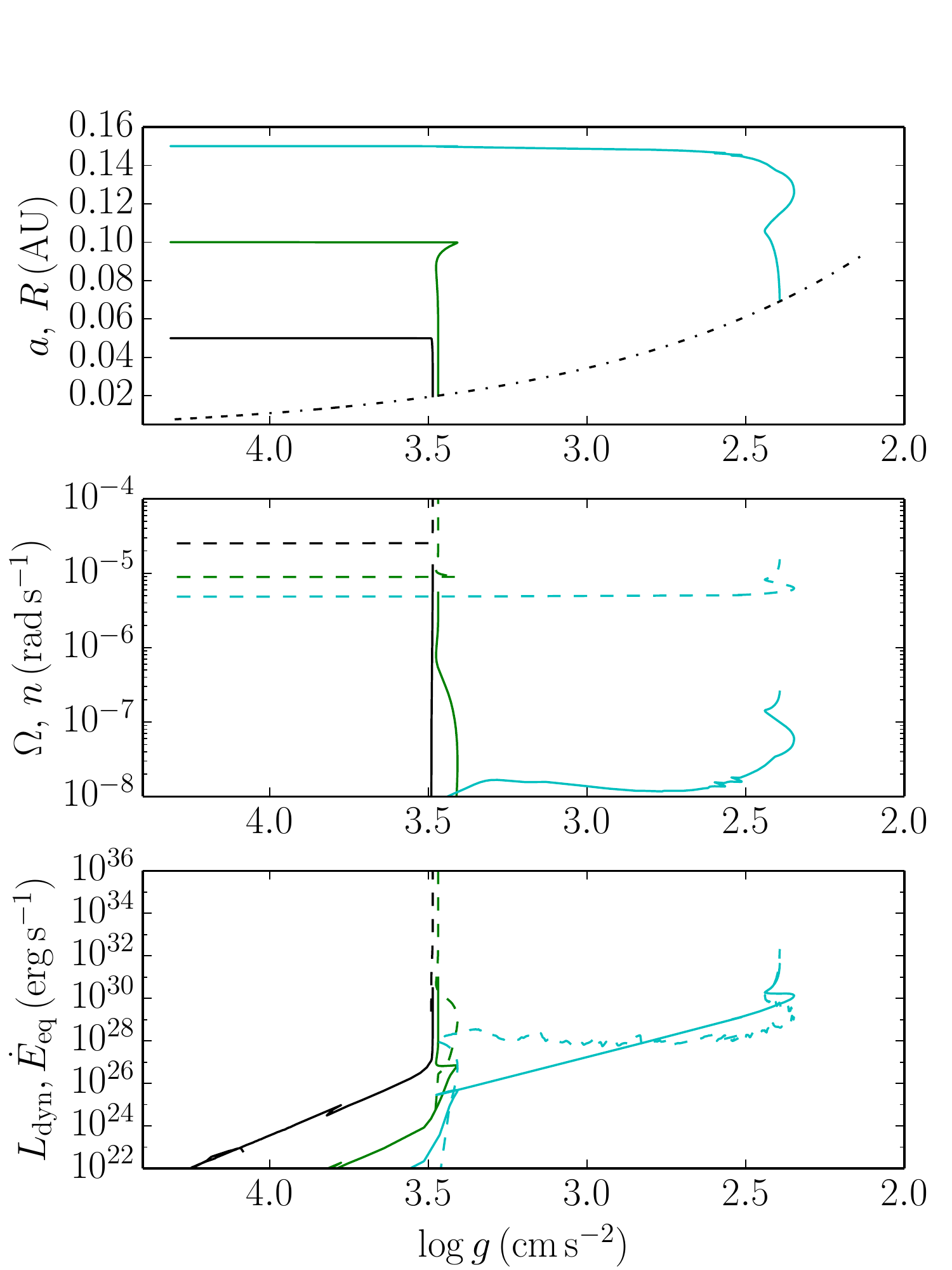}
	\caption{Same as Figure \ref{2M_10MJ_ZN.pdf} but with GN's viscosity prescription. }
	\label{2M_10MJ_GN.pdf}
\end{figure}

Figure \ref{2M_10MJ_GN.pdf} is the same as Figure \ref{2M_10MJ_ZN.pdf} but uses GN's viscosity prescription. There is almost no difference for the merger time for the case of $a_{\rm init}=$ 0.05 and 0.1 AU since dynamical tides dominate. However, since the $a_{\rm init}=$ 0.15 AU case is dominated by the equilibrium tide, it merges later than for Zahn's prescription, since $\dot{E}_{\rm eq}$ is much smaller. For the $a_{\rm init}=$ 0.15 AU case, the system merges at $\log\,g=2.4$, which is near dredge-up, where the convective envelope reaches the hydrogen burning shell, and the radius first contracts and then expands.

\begin{figure}
	\centering
	\includegraphics[width=0.5\textwidth]{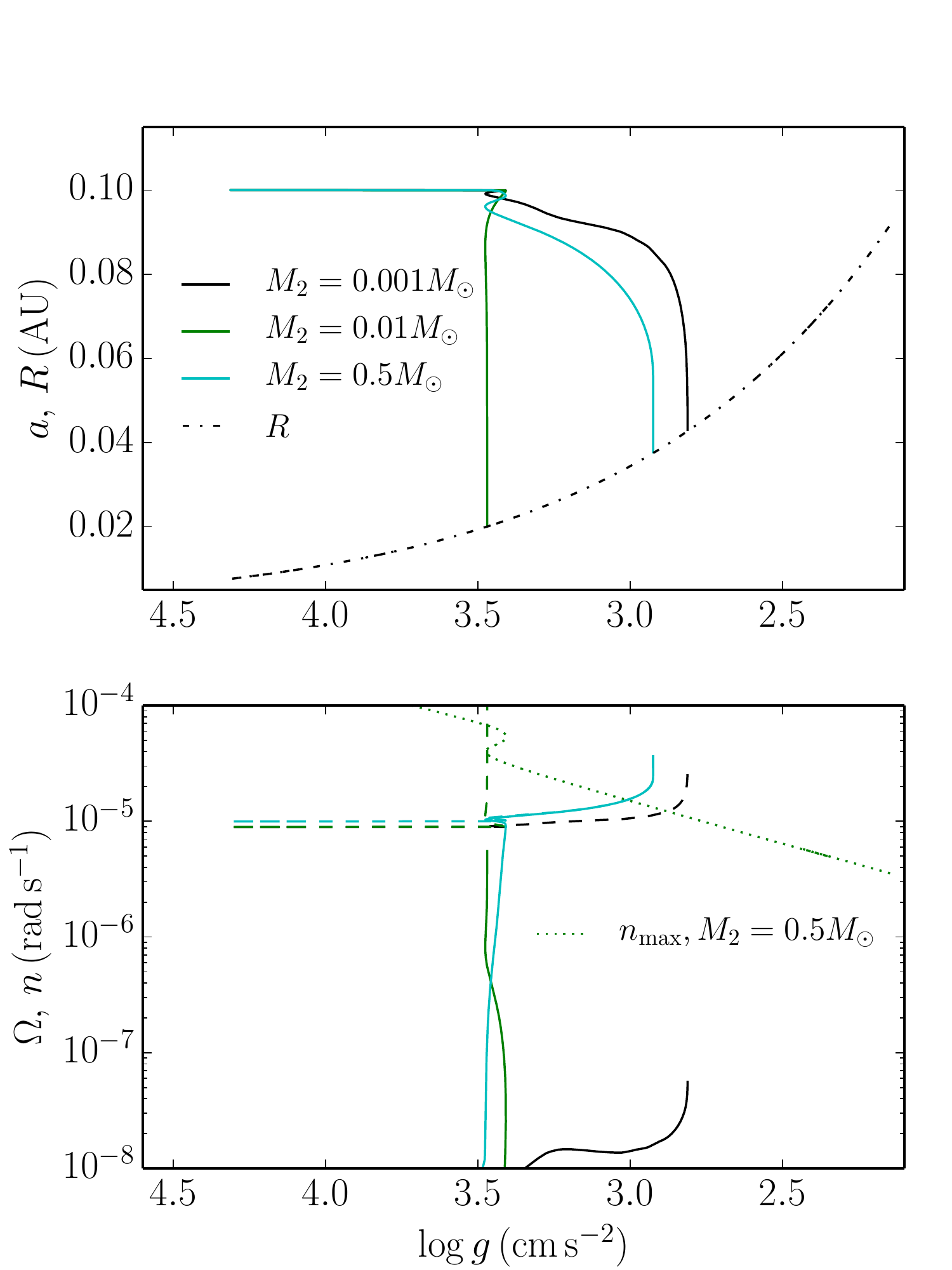}
	\caption{Same as Figure \ref{1M_015_big_companion_ZN.pdf} but with $M_1=2\, M_\odot$, and $M_2=$ 0.001 (black), 0.01 (green) and 0.5 (cyan), starting separation $a_{\rm init}=0.1\, \rm AU$, and uses Zahn's prescription for viscosity. In the lower panel, the green dotted line displays the orbital frequency above which the Darwin instability occurs ($n_{\rm max}$), evaluated with $M_2=0.5M_{\odot}$.} 
	\label{2M_010_big_companion_ZN.pdf}
\end{figure}

Figure \ref{2M_010_big_companion_ZN.pdf} again uses $M_1=2\, M_\odot$ and compares tracks with companion masses $M_2=0.001, 0.01$ and $0.5\,M_{\odot}$ for $a_{\rm init}=0.1\, \rm AU$ with Zahn's prescription for viscosity. In the second panel, the z-shape in $n_{\rm max}$ near $\log\,g=3.5 \sim 3.4$ occurs between central hydrogen exhaustion and shell ignition, where the star first shrinks and then expands. Similar to the $M_1=1\, M_\odot$ case, the two low mass cases are not synchronized while the higher mass case is synchronized until the Darwin instability. 

\begin{figure}
	\centering
	\includegraphics[width=0.5\textwidth]{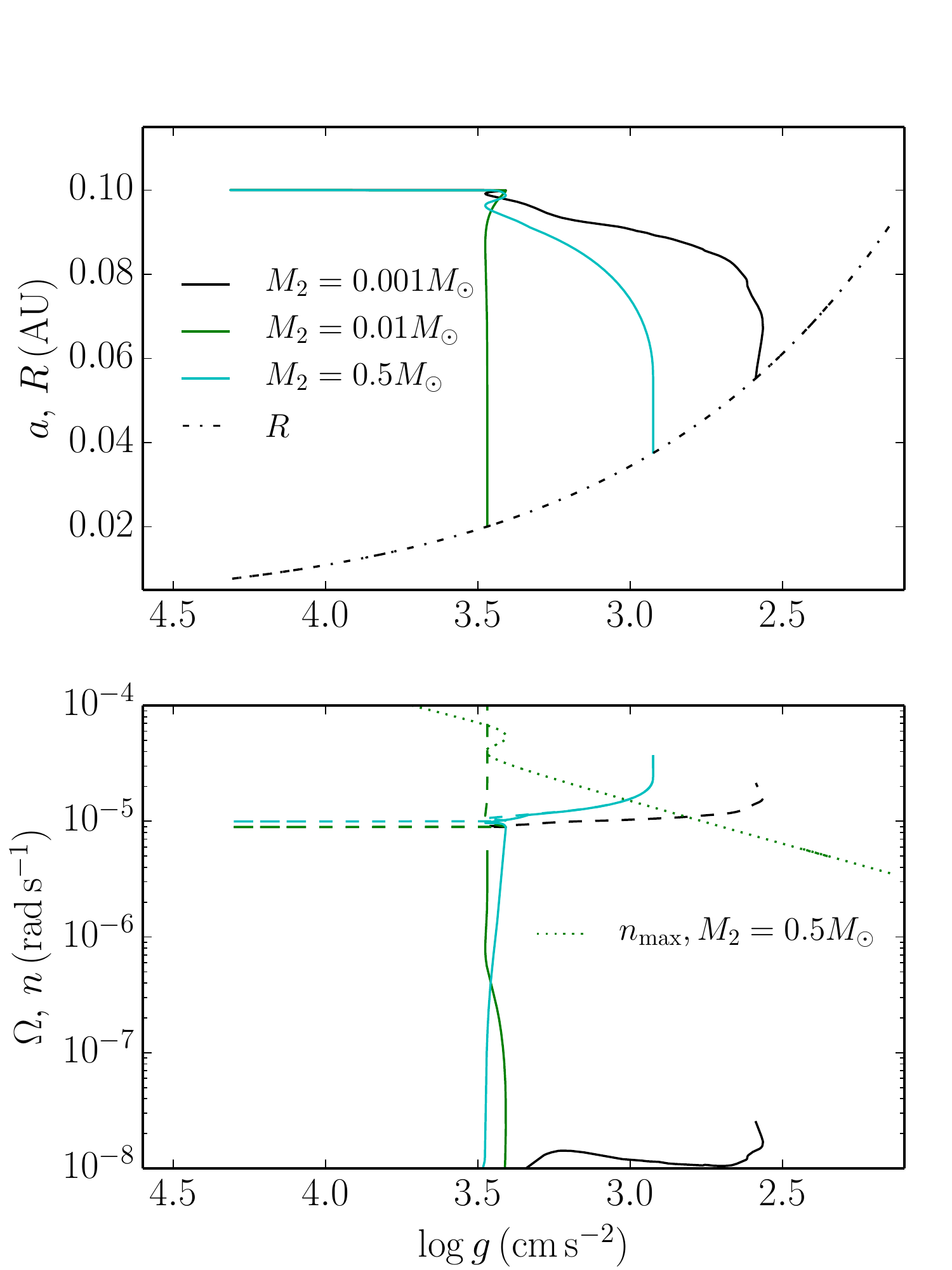}
	\caption{Same as Figure \ref{2M_010_big_companion_ZN.pdf} but with GN's viscosity prescription. In the lower panel, the green dotted line displays the orbital frequencies above which the Darwin instability occurs ($n_{\rm max}$), evaluated with $M_2=0.5M_{\odot}$.} 
	\label{2M_010_big_companion_GN.pdf}
\end{figure}

Figure \ref{2M_010_big_companion_GN.pdf}  is the same as Figure \ref{2M_010_big_companion_ZN.pdf}, but uses GN's viscosity prescription. For the $M_2=0.001M_{\odot}$ case, dynamical tides are more important than for Zahn's viscosity, and the merger occurs slightly later compared to Figure \ref{2M_010_big_companion_ZN.pdf}.
For the $M_2=0.01M_{\odot}$ case, the spin is not synchronized and dynamical tides dominate, and so the merger is near that in Figure \ref{2M_010_big_companion_ZN.pdf}. For the $M_2=0.5M_{\odot}$ case, the spin synchronizes leading to a merger at a similar time as in Figure \ref{2M_010_big_companion_ZN.pdf}.



\subsection{3 $M_{\odot}$ Model}

The $M_1=3\, M_{\odot}$ case is qualitatively similar to that of $M_1=2\, M_{\odot}$. Tidal dissipation is suppressed during the MS, effectively turning on when the convective envelope thickens near $\log g \simeq 3.0$ on the SGB.

\begin{figure}
\centering
\includegraphics[width=0.5\textwidth]{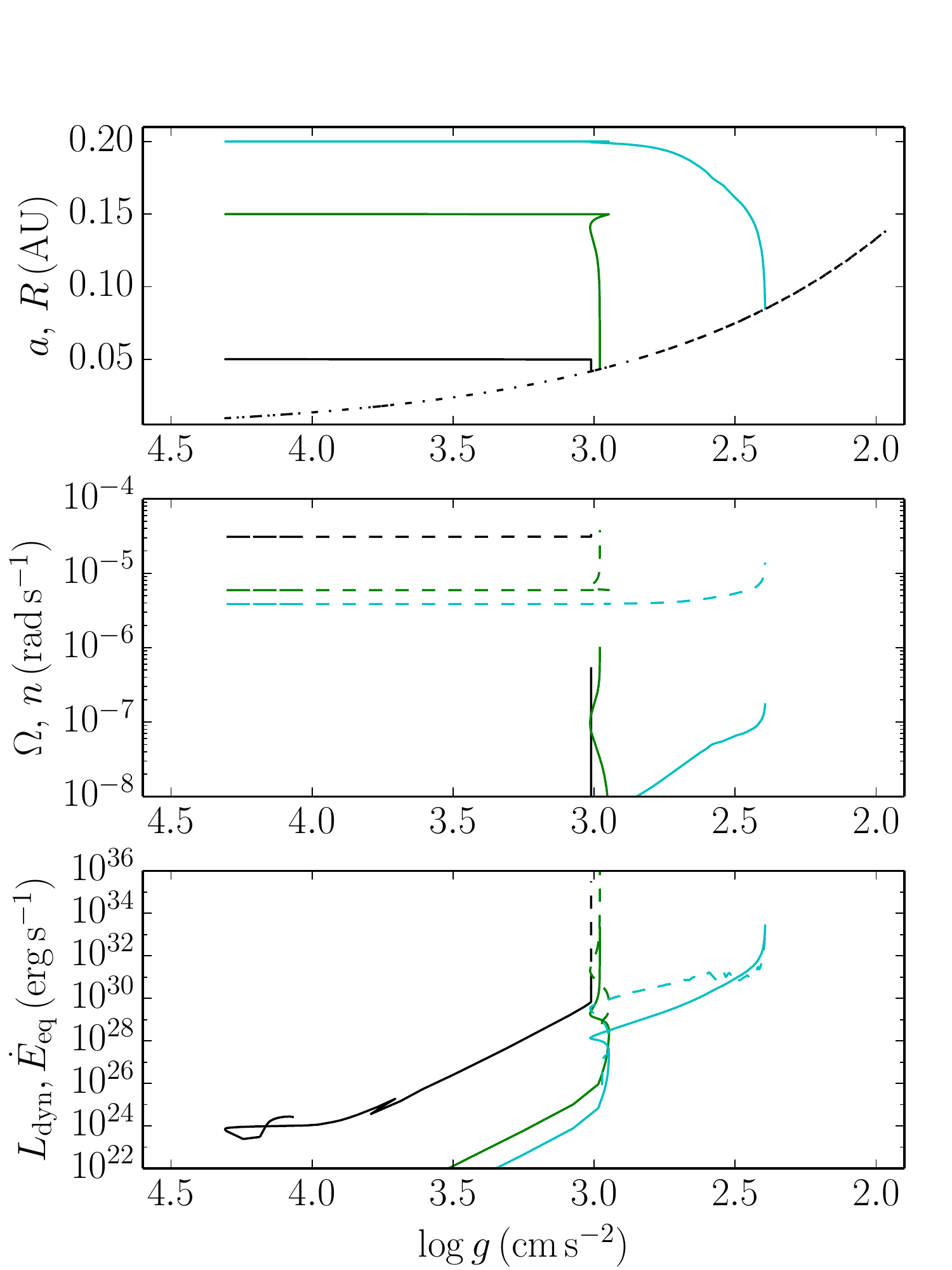}
\caption{Same as Figure \ref{1M_10MJ_ZN} but with $M_1=3\, M_\odot$, $M_2=0.01\, M_\odot$ and $a_{\rm init}=0.05$ (black lines), 0.15 (green lines) and 0.2 (cyan lines) AU. Zahn's prescription is used.}
\label{3M_10MJ_ZN.pdf}
\end{figure}
 
The bottom panel of Figure \ref{3M_10MJ_ZN.pdf} shows that the dynamical tide is much bigger than equilibrium tide when it turns on, and immediately causes both the $a=0.05$ and $0.15\, \rm AU$ orbits to decay. Only the $a=0.2\, \rm AU$ orbit is sufficiently wide that the star has time to move up the RGB and the equilibrium tide can dominate. None of the cases have synchronous spin for this companion mass (middle panel).
 
\begin{figure}
\centering
\includegraphics[width=0.5\textwidth]{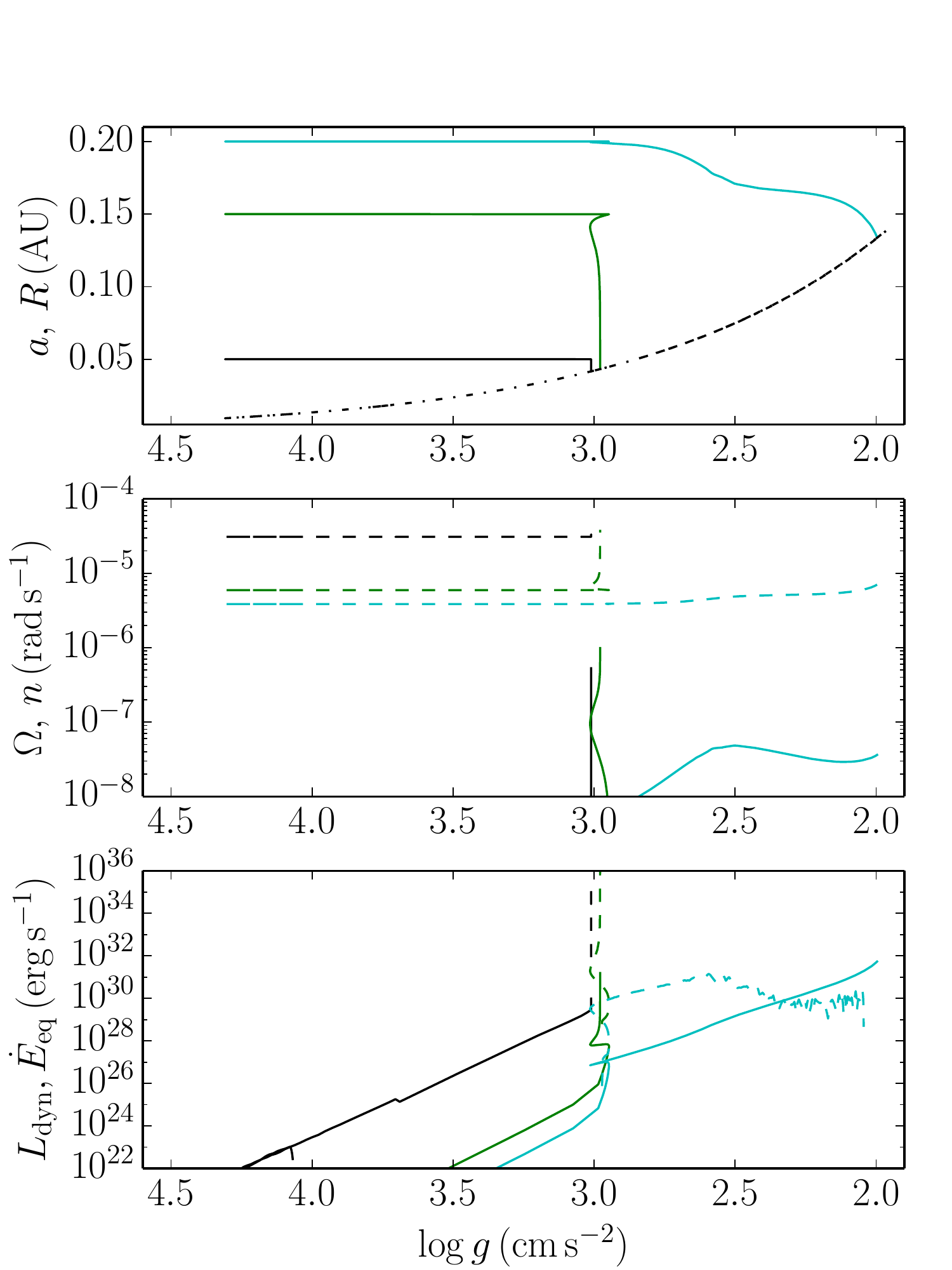}
\caption{Same as Figure \ref{3M_10MJ_ZN.pdf} but with GN's viscosity prescription.}
\label{3M_10MJ_GN.pdf}
\end{figure}

Figure \ref{3M_10MJ_GN.pdf} shows the same experiment as Figure \ref{3M_10MJ_ZN.pdf} but with GN's viscosity. The merger in the $a_{\rm init}=0.05$ and $0.15$ AU cases is due to the dynamical tide and agrees closely with Figure \ref{3M_10MJ_ZN.pdf}. The $a_{\rm init}=0.2$ AU case ends up dominated by the equilibrium tide, and is not synchronized, and hence merges much later than that in Figure \ref{3M_10MJ_ZN.pdf}.

\begin{figure}
\centering
\includegraphics[width=0.5\textwidth]{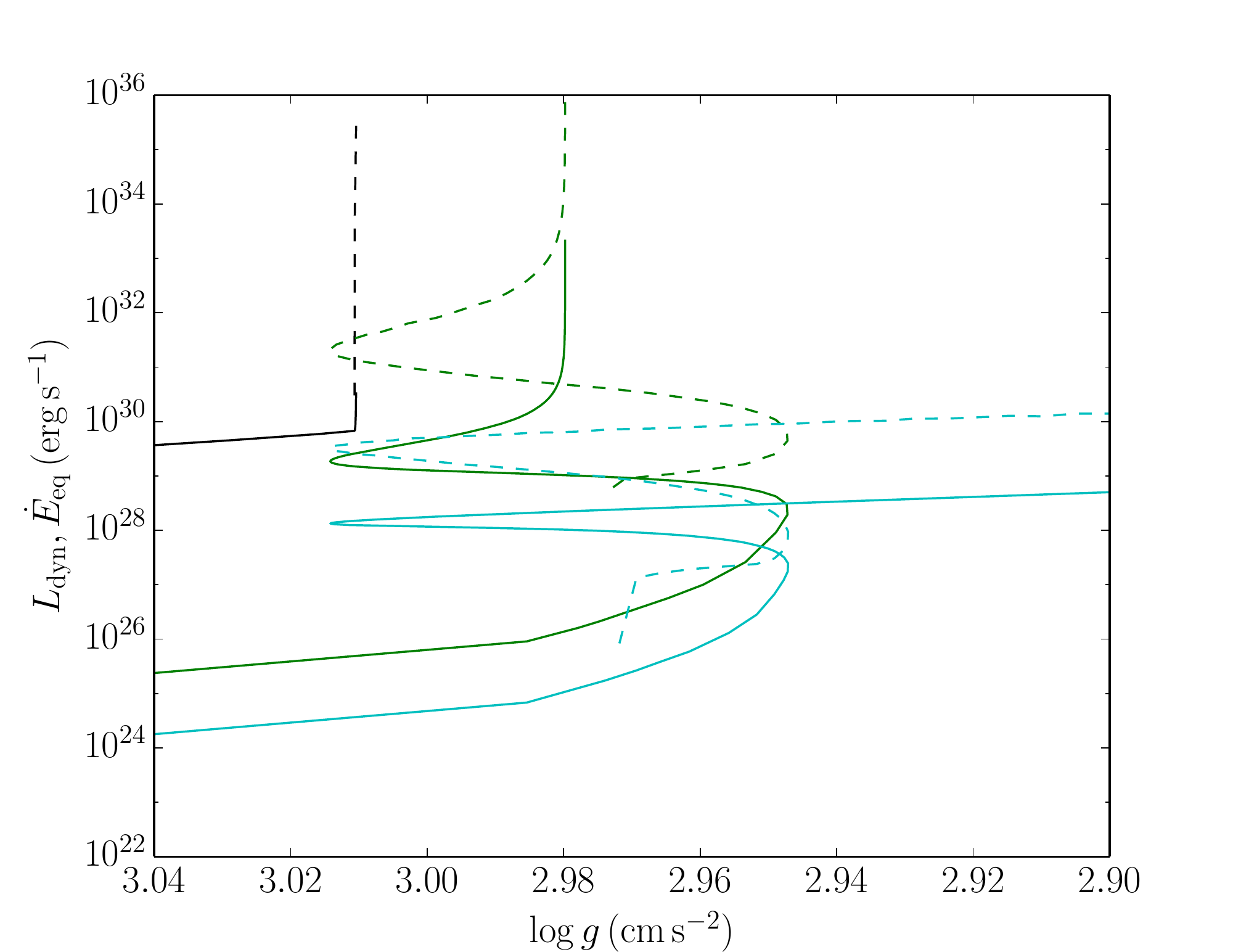}
\caption{Close-up of the bottom panel of Figure \ref{3M_10MJ_ZN.pdf}, near where the convective envelope deepens and tidal friction increases dramatically.}
\label{3M_10MJ_ZN_close.pdf}
\end{figure}

Figure \ref{3M_10MJ_ZN_close.pdf} shows a close-up of the dissipation rates in the bottom panel of Figure \ref{3M_10MJ_ZN.pdf}. For all three cases, the dynamical tide (dashed lines) dominates the equilibrium tide (solid lines) as soon as it turns on. 

\begin{figure}
	\centering
	\includegraphics[width=0.5\textwidth]{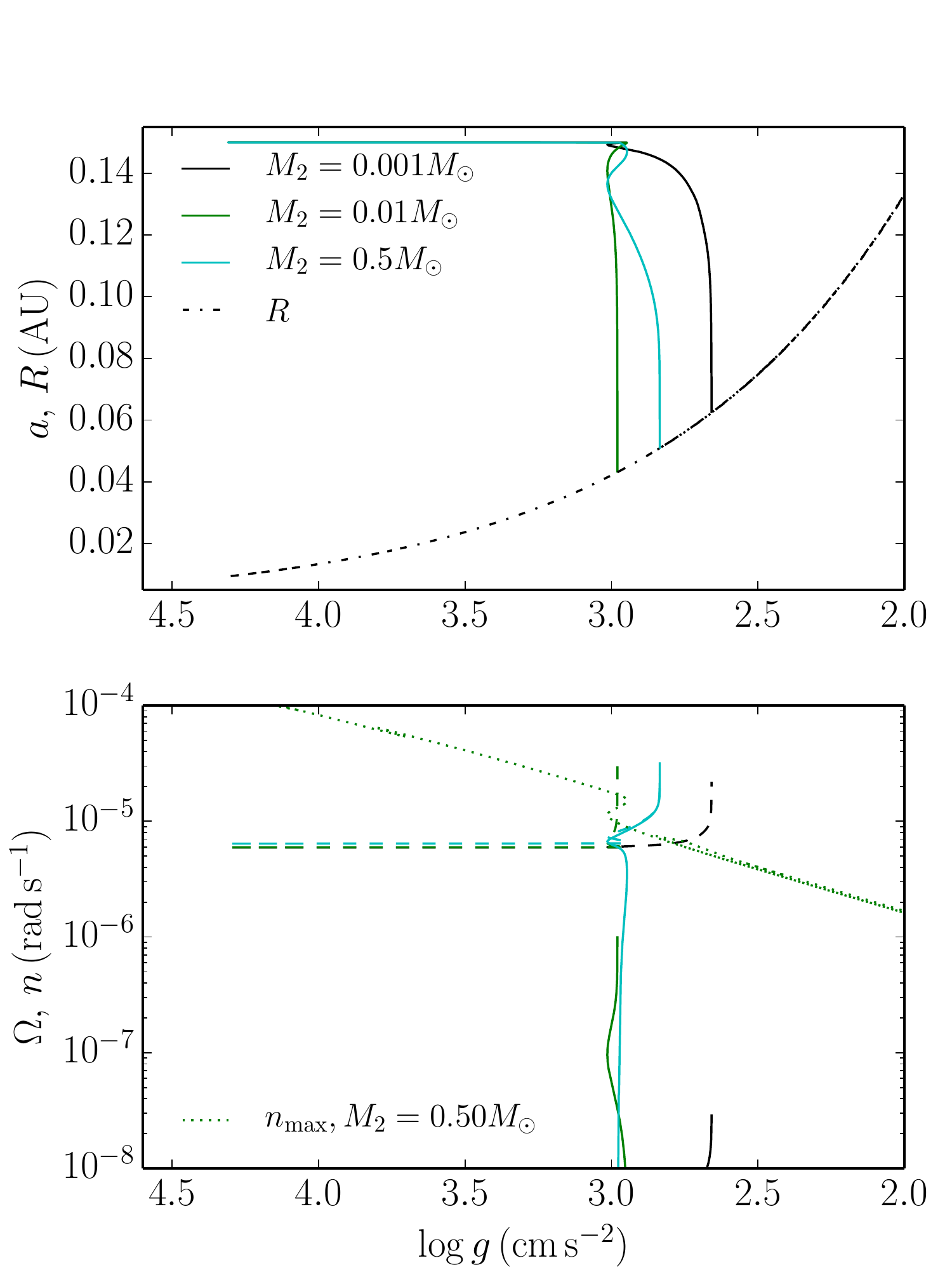}
	\caption{Same as Figure \ref{1M_015_big_companion_ZN.pdf} but with $M_1=3\, M_\odot$, and $M_2=$ 0.001 (green), 0.01 (blue) and 0.5 (red), and starting separation $a_{\rm init}=0.15\, \rm AU$. In the lower panel, the green dotted line displays the orbital frequencies above which the Darwin instability occurs ($n_{\rm max}$), evaluated with $M_2=0.5M_{\odot}$.}
	\label{3M_015_big_companion_ZN.pdf}
\end{figure}

Figure \ref{3M_015_big_companion_ZN.pdf} again compares the evolution with Zahn's viscosity and including the dynamical tide for three different companion masses. All three cases merge over a small range of $\log g$ soon after the dynamical tide turns on. For the most massive $M_2=0.5\, M_\odot$ companion, in spite of the fact that the spin becomes synchronous, the binary is still short lived since the Darwin instability turns on at roughly the same time.

\begin{figure}
	\centering
	\includegraphics[width=0.5\textwidth]{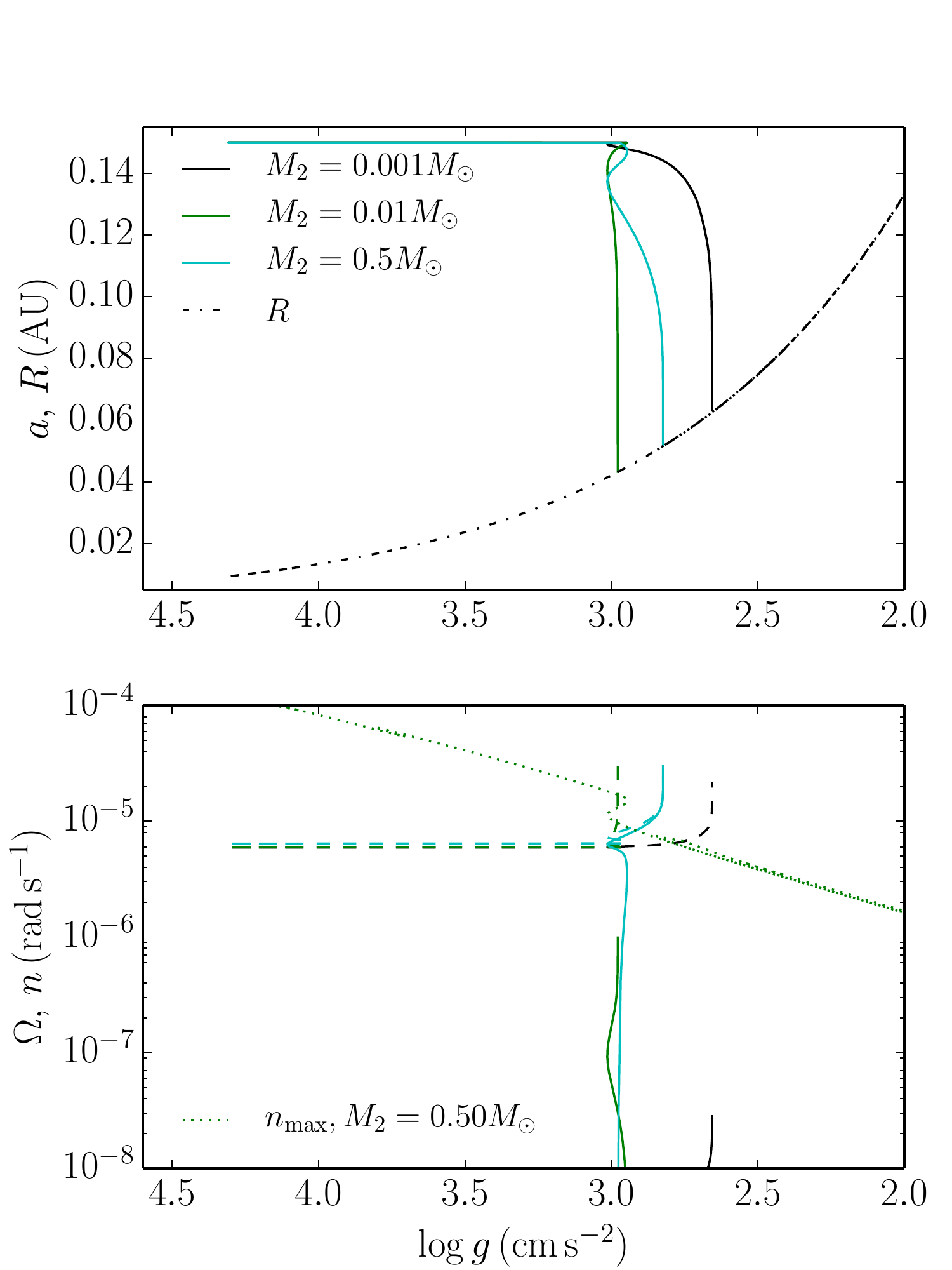}
	\caption{Same as Figure \ref{3M_015_big_companion_ZN.pdf} but with GN's viscosity prescription.}
	\label{3M_015_big_companion_GN.pdf}
\end{figure}

The result for GN's viscosity in Figure \ref{3M_015_big_companion_GN.pdf} is very similar to Figure \ref{3M_015_big_companion_ZN.pdf}. As the dynamical tide for the $3M_{\odot}$ case is strong for all the three companion masses, the different viscous dissipation reduction factors make almost no change in the orbital evolution.



\section{Critical Semi-major Axis for Mergers}
\label{sec:acrit}

Examples of orbital decay were shown in Section \ref{sec:examples} to understand the importance of the turbulent viscosity prescription, the strength of dynamical versus equilibrium tide, and synchronous spin at each stage in a star's evolution. In this section, calculations of the ``critical radius", $a_{\rm crit}(t)$, which depends on the age of the system, are presented. The critical radius is defined as the {\it initial} separation out to which orbits would have decayed down to the surface of the primary by the age $t$. 
Few binaries are expected to be found with (present day) $a \la a_{\rm crit}(t)$ since those orbits should have decayed and the binary already merged, while binaries with $a \ga a_{\rm crit}$ are relatively unaffected by orbital decay. Hence if a lack of systems is observed for some range of semi-major axis, the plot of $a_{\rm crit}$ versus $\log g$ shows which range might be absent of binaries due to tides, and for which range the lack of systems must have some other explanation as tides become ineffective. 

\begin{figure}
	\centering
	\includegraphics[width=0.5\textwidth]{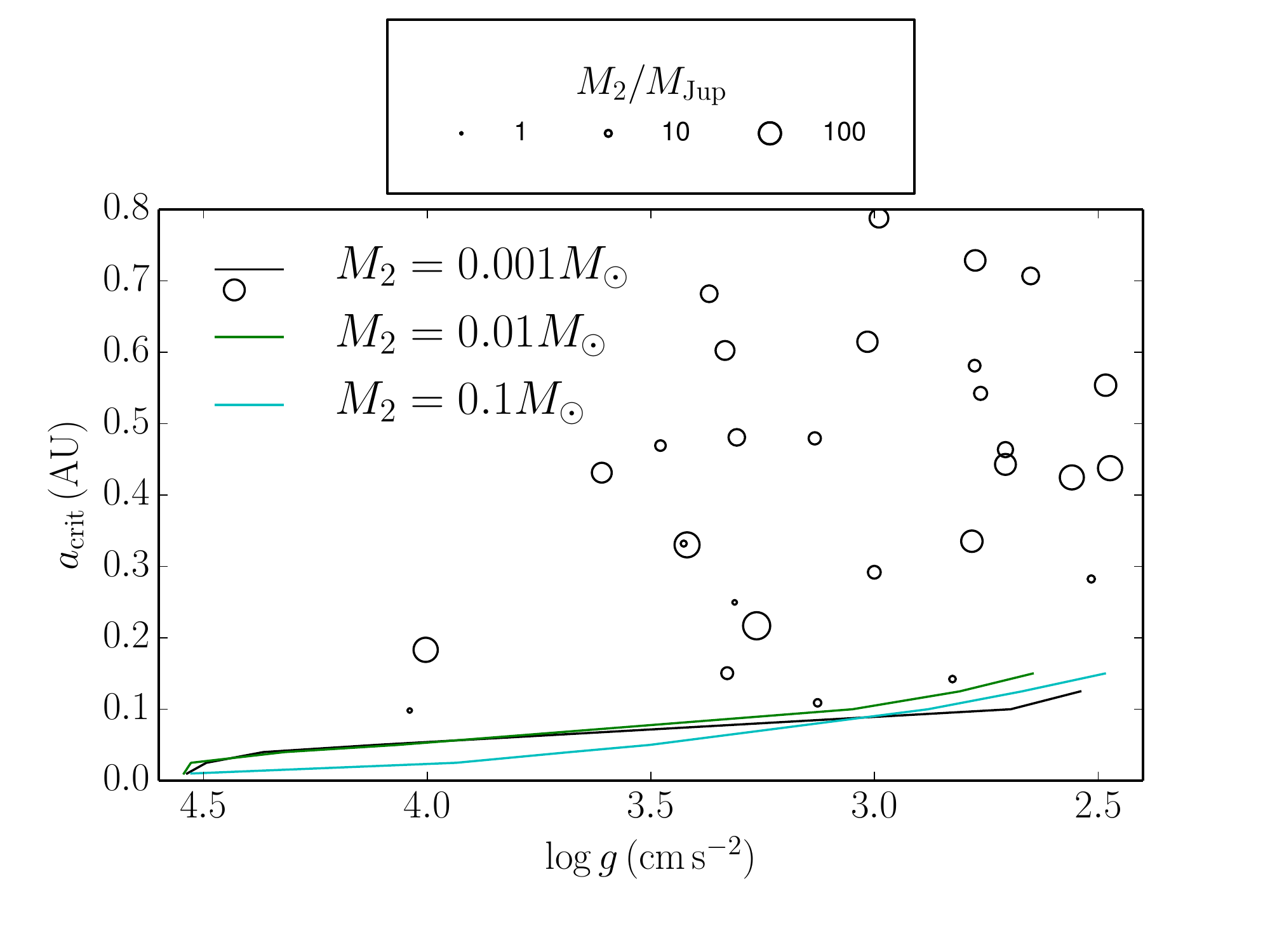}
	\caption{The critical semi-major axis $a_{\rm crit}$ versus $\log\,g$ for $M_1=1M_{\odot}$ with three companion masses: $M_2 = 0.001M_{\odot}$ (black), $M_2 = 0.01M_{\odot}$ (green) and $M_2 = 0.1M_{\odot}$ (cyan). The black dots are the APOGEE data \citep{2016AJ....151...85T}, with $M_1$ between $0.5M_{\odot}$ and $1.5M_{\odot}$, and $M_2$ between $1M_{\rm Jup}$ - $100M_{\rm Jup}$. }
	\label{acrit_1Msun.pdf}
\end{figure}

\begin{figure}
	\centering
	\includegraphics[width=0.5\textwidth]{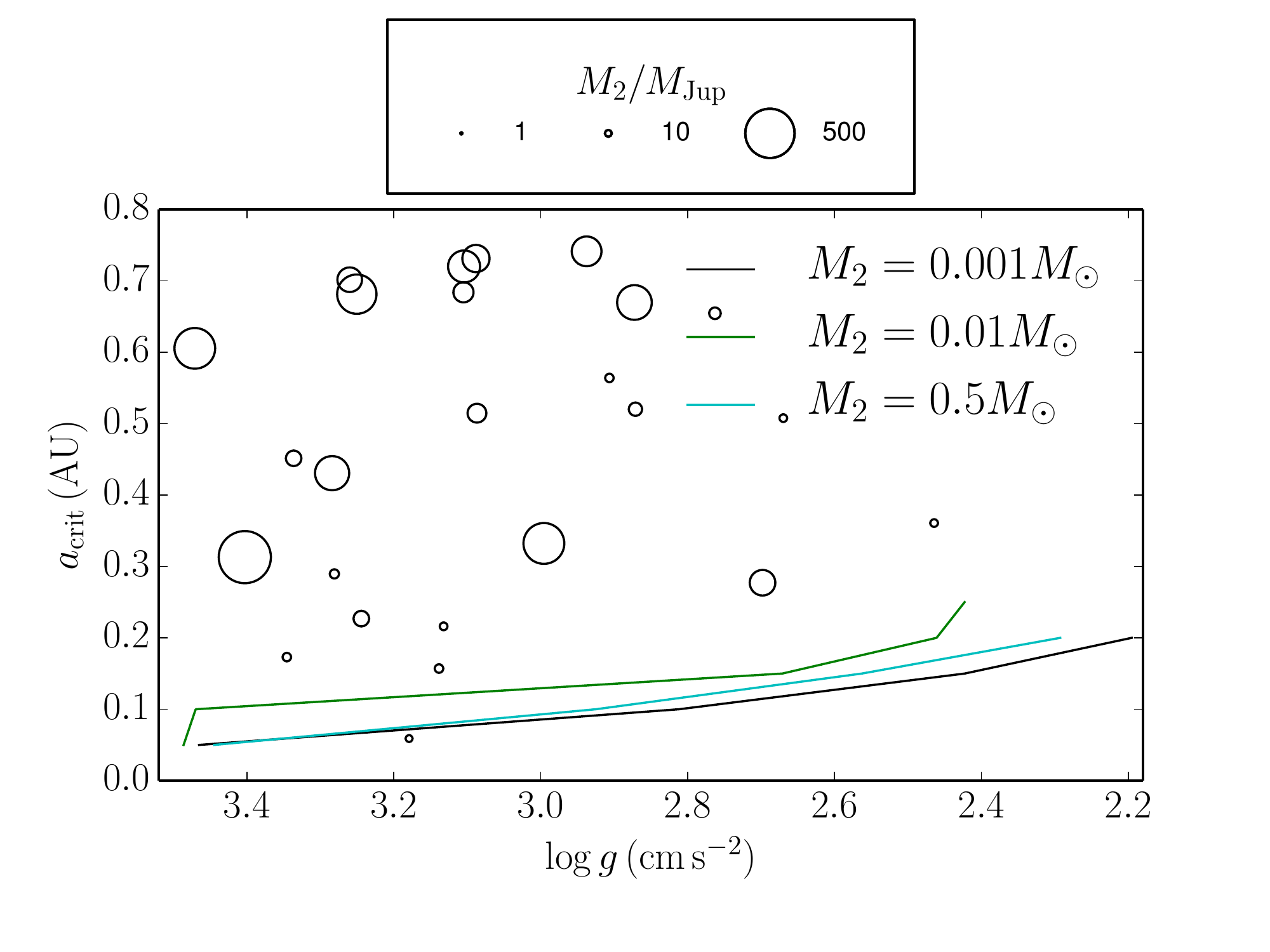}
	\caption{Same as Figure \ref{acrit_1Msun.pdf} for $M_1=2M_{\odot}$ and $M_2 = 0.001$ (black), $0.01$ (green) and $0.5M_{\odot}$ (cyan). The black dots are the APOGEE data \citep{2016AJ....151...85T}, with $M_1=1.5$ - $2.5M_{\odot}$ and $M_2=1$ - $500M_{\rm Jup}$.}
	\label{acrit_2Msun.pdf}
\end{figure}

\begin{figure}
	\centering
	\includegraphics[width=0.5\textwidth]{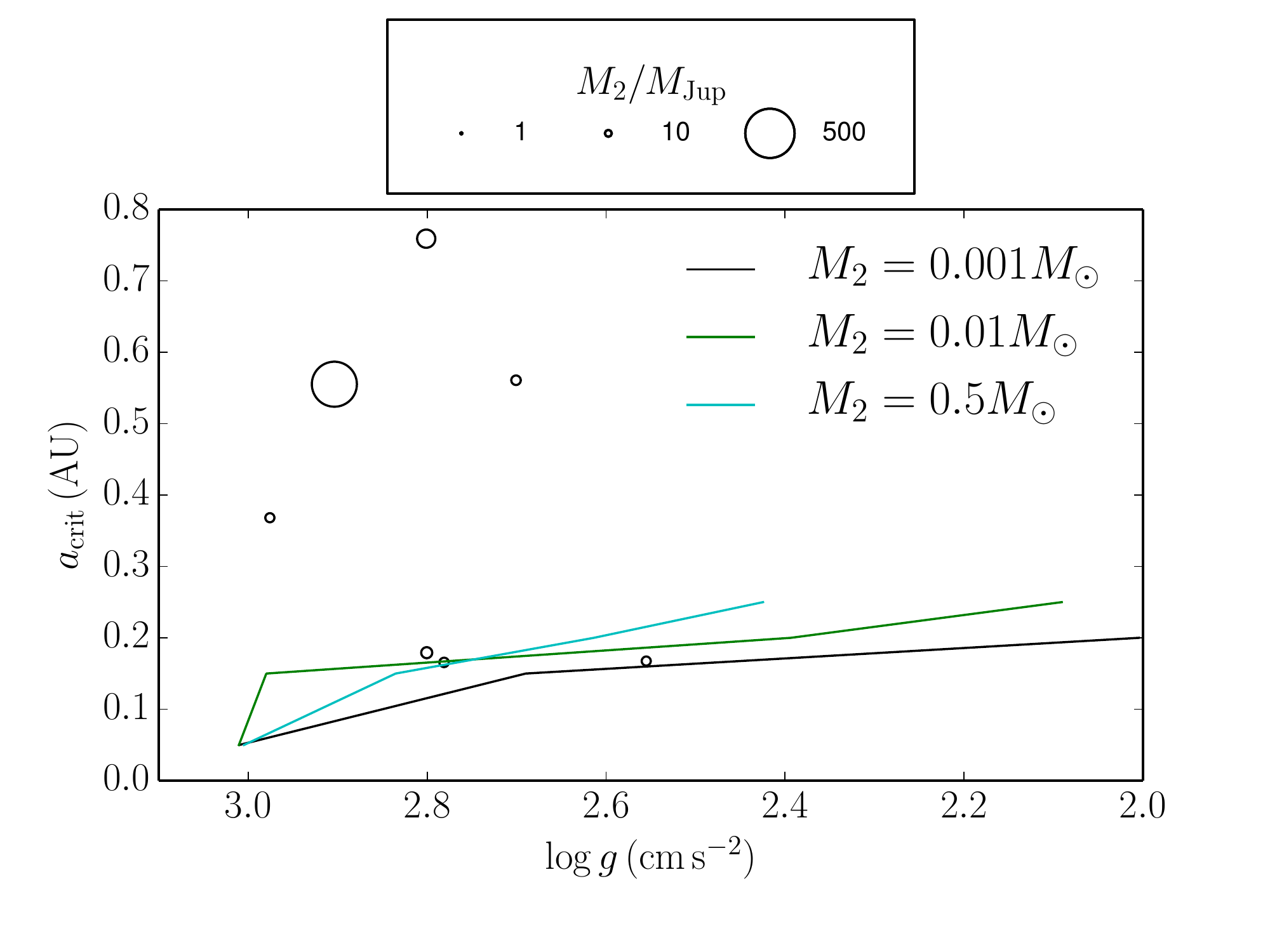}
	\caption{Same as Figure \ref{acrit_1Msun.pdf}  for $M_1=3M_{\odot}$ with three companion masses $M_2 = 0.001M_{\odot}$ (black), $M_2 = 0.01M_{\odot}$ (green) and $M_2 = 0.5M_{\odot}$ (cyan). The black dots are the APOGEE data \citep{2016AJ....151...85T}, with $M_1=2.5$ - $3.5M_{\odot}$ and $M_2=1$ - $500M_{\rm Jup}$.}
	\label{acrit_3Msun.pdf}
\end{figure}

Figures \ref{acrit_1Msun.pdf}, \ref{acrit_2Msun.pdf} and \ref{acrit_3Msun.pdf} show our calculations of $a_{\rm crit}$ versus $\log g$ for a primary mass of $M_1=1,2,3\, M_\odot$, respectively, with the open circles showing data for binary systems from the APOGEE survey. The different lines in each plot are for different secondary masses, $M_2$.  Zahn's prescription for reduced viscosity will be used in all plots.
This reduces to standard (un-reduced) viscosity when eddy turnover times are short. 


Figure \ref{acrit_1Msun.pdf} compares calculations of $a_{\rm crit}$ versus $\log\,g$ for $M_1=1M_{\odot}$. These calculations are compared to APOGEE binaries \citep{2016AJ....151...85T} for SGB and RGB primary stars in the mass range $M_1=0.8$ - $1.5\, M_\odot$. 
All the observed systems have $a>a_{\rm crit}$, meaning that the orbital decay rate is small compared to the stellar evolution timescale. The lack of systems at $a<a_{\rm crit}$ may be interpreted as either the population of closer systems with $a < a_{\rm crit}$ have already merged due to orbital decay, or such close binary systems are rare or never formed in the first place.
Note that for extremely small $a_{\rm init}$, companions of mass $M_2 = 0.001$ - $0.1\, M_{\odot}$ would suffer orbital decay before the end of the MS, and well before the RGB. This corresponds to the starting point of $a_{\rm init}$ lines in the bottom left of the figure.

Figure \ref{acrit_2Msun.pdf} shows the $M_1=2M_{\odot}$ case. Due to the thin surface convection zone, and a central convection zone, dynamical tides are assumed ineffective on the MS. The surface convection zone deepens near $\log g \simeq 3.5$, at which point both dynamical and equilibrium tides increase dramatically. This causes rapid orbital decay over a range of small orbital separation. At $a_{\rm crit}=0.05$ AU, dynamical tides rapidly shrink the orbit, and even the high mass companions can't synchronize the orbit within the short orbital decay timescale. Therefore the $a_{\rm crit}$ lines for the three $M_2$ are close. At $a_{\rm crit} = 0.1$ AU for the $M_2 = 0.01M_{\odot}$ secondary, dynamical tides are still strong and the orbit shrinks at $\log\,g=3.5$. Slow orbital decay occurs both for the high mass companion, due to synchronous spin, and low mass companions due to the weak tidal force. All but one of the APOGEE systems have $a \geq a_{\rm crit}$, again showing that the $a \leq a_{\rm crit}$ systems, especially with $\log g \la 3.5$, may have already been destroyed.

Figure \ref{acrit_3Msun.pdf} shows the $M_1=3M_{\odot}$ case. Similar to the $2M_{\odot}$, $a_{\rm crit}=0.05$ AU case, dynamical tides are strong and synchronization does not occur. The system separation decreases quickly at $\log\,g=3.0$ for the range $M_2=0.001$ - $0.5M_{\odot}$. For the $M_2 = 0.5M_{\odot}$ secondary, the distance of 0.2 AU is far enough away for the high mass companion to synchronize the orbit, which makes the system survive longer compared with using a $M_2 = 0.01M_{\odot}$ or $M_2 = 0.001M_{\odot}$ secondary. The observed systems are again shown in black circles. In this case there are three observed binaries near the $a=a_{\rm crit}$ lines, that may be undergoing more rapid orbital decay. Wide orbits with with $a_{\rm obs}>0.8$ AU are not shown in Figure \ref{acrit_1Msun.pdf}, \ref{acrit_2Msun.pdf} and \ref{acrit_3Msun.pdf}.


Analytic estimates for $a_{\rm crit}$ are given in Appendix \ref{sec:analytic_acrit} in Equations \ref{eq:aD_vs_R}, \ref{eq:acritstd}, \ref{eq:acritz}, \ref{eq:acritgn} and \ref{acrit_dyn_R}. Since $\dot{E}$ scales as a high power of $a$, $a_{\rm crit}$ is a very weak function of $M_2$ for equilibrium and dynamical tides with non-synchronized primary. The dependence on $R_1$ is nearly linear for the equilibrium tide, but is dominated by the SGB and base of the RGB for the dynamical tide. 

\begin{figure}
	\centering
	\includegraphics[width=0.5\textwidth]{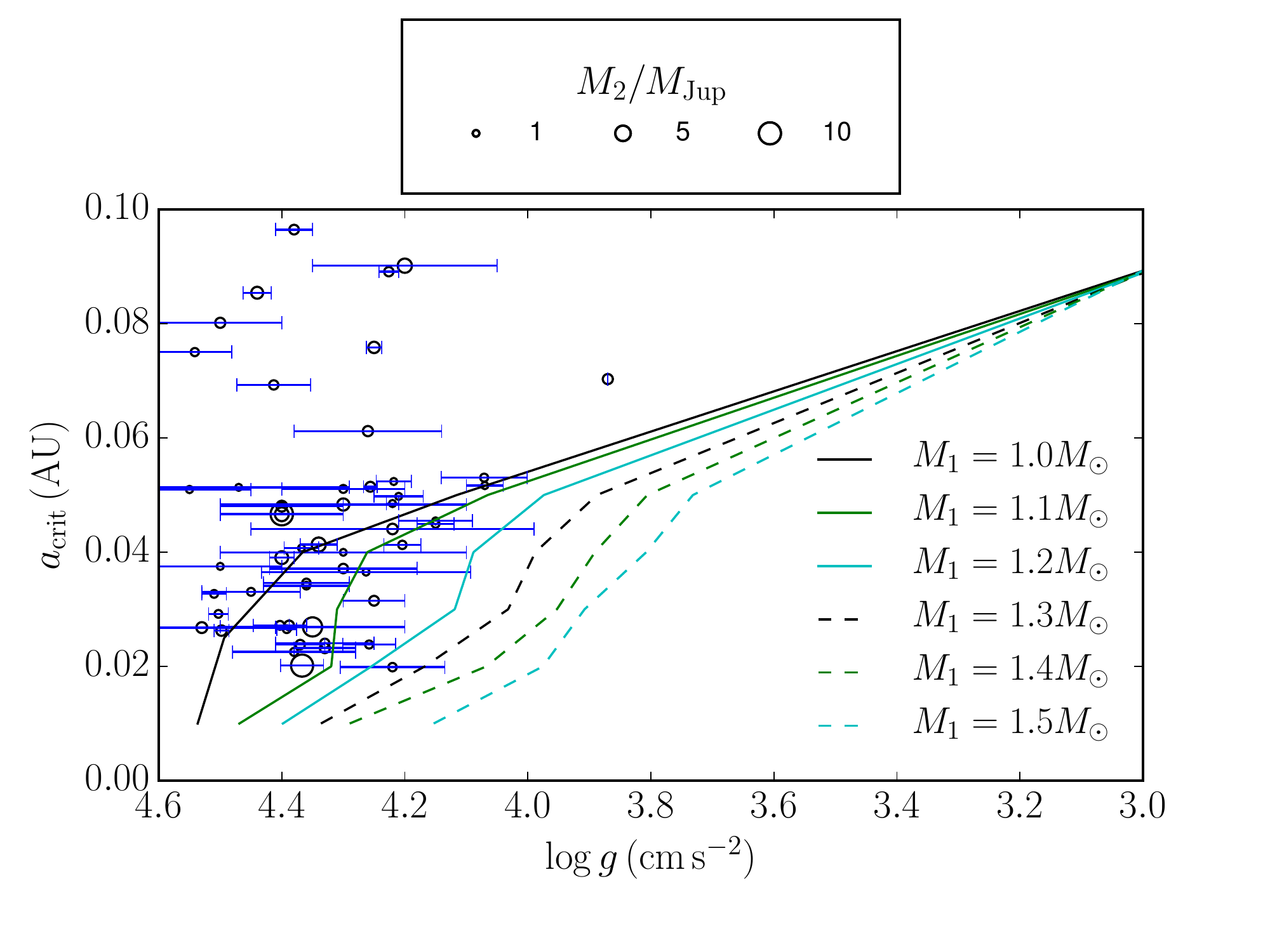}
	\caption{Critical semi-major axis $a_{\rm crit}$ versus $\log\,g$ for companion mass $M_2=10^{-3}\, M_\odot$ and primary masses $M_1=1.0M_\odot$ (blue), $1.1M_\odot$ (black), $1.2M_\odot$ (magenta), $1.3M_\odot$ (yellow), $1.4M_\odot$ (orange), $1.5M_\odot$ (cyan). The exoplanet plus host star data from \citet{2014PASP..126..827H} is given by open circles with error bars for $\log\,g$ in blue. }
	\label{acrit_krxir_1015Msun_exo.pdf}
\end{figure}

Next, the calculations of $a_{\rm crit}$ are  compared to binaries containing an exoplanet and host star from \citet{2014PASP..126..827H} (downloaded from exoplanets.org). Figure \ref{acrit_krxir_1015Msun_exo.pdf} shows stars in the mass range $M_1=1-1.3\, M_{\odot}$ with planetary mass companions. These stars are in the MS or early SGB. \footnote{The reason why the upper limit is set at $1.3M_\odot$ is because stars with $M>1.3M_\odot$ have a convective core and thin convective envelope, hence both the dynamical and equilibrium tide dissipation rates are expected to be smaller by comparison, giving small $a_{\rm crit}$ during the MS stage.}

Immediately apparent in Figure \ref{acrit_krxir_1015Msun_exo.pdf} is the strong dependence of $a_{\rm crit}$ on stellar mass over the gravity range $\log g = 4.0-4.5$ and mass range $M_1=1.0-1.5\, M_\odot$. The sharp decrease in $a_{\rm crit}$ at fixed $\log g$ for slightly higher $M_1$ is due to the dynamical tide shutting off for stars with convective cores. The weaker equilibrium tide dissipation rate leads to much smaller $a_{\rm crit}$ when convective cores are present. Stars with a convective core on the MS initially have small $a_{\rm crit}$ while the convective core is present, and then $a_{\rm crit}$ suddenly increases at the end of the MS when the core becomes radiative and the convective envelope deepens. In addition, for the case of $M_1=1.2M_{\odot}$ - $1.5M_{\odot}$, at nearly the same time that the central convective core ceases, $k_r\xi_r$ becomes close to 1 and the dynamical tides become effective. 
The reason of the increase in $k_r\xi_r$ is that the density near the star center increases and the inner turning point of the wave moves inward as composition gradients also increase during the evolution. This result agrees with \cite{2010MNRAS.404.1849B}.

This strong dependence on stellar mass implies that care must be taken when comparing an observed orbital separation, $a$, with a calculated critical separation, $a_{\rm crit}$. Uncertainties in spectroscopic determination of stellar mass and radius may not be small enough to accurately decide which theory curve is appropriate. The error bars on primary star gravity in Figure \ref{acrit_krxir_1015Msun_exo.pdf} are $\Delta(\log g) \simeq 0.1-0.2$, and uncertainties in fitting mass are at the level $\Delta(M_1) \simeq 0.1-0.2\, M_\odot$. If, for example, all data points in the range $\log g = 4.2 - 4.5$ were compared with the $M_1=1.0\, M_\odot$ line, a large number of systems would have $a \la a_{\rm crit}$ since dynamical tide dissipation is present for that stellar mass. However, if the stellar mass was slightly higher, with $M_1=1.1-1.2\, M_\odot$, then most of the points have $a>a_{\rm crit}$. A detailed comparison of each system with the measured $M_1$ indeed shows that nearly all exoplanets plus host star have $a \ga a_{\rm crit}$. However, given the size of the error bars as compared to the rapid change in the theory curves this result should be taken with some caution.

For non-transiting binaries detected by the radial velocity method, only the minimum mass $M_2 \sin(i)$ is measured, where $i$ is the binary inclination. If the primary is not synchronized, using $M_2\sin(i)$ as the secondary mass instead of $M_2$ will lead to smaller $a_{\rm crit}$, although this dependence is very weak (see Appendix \ref{sec:analytic_acrit}). This would tend to make points have larger $a/a_{\rm crit}$. Next, if the primary star is synchronized, $a_{\rm D}$ varies inversely with $M_2$, tending to make the $a_{\rm crit}$ larger, and would tend to make smaller $a/a_{\rm crit}$. In practice systems with $\sin(i) \ll 1$ are rare.


For the APOGEE systems and the exoplanet host stars, our results are consistent with nearly all systems having $a \ga a_{\rm crit}$. A natural question is what the distribution of orbital separation would look like if a large number of systems did indeed have $a \ll a_{\rm crit}$.
This would imply that a large number of binaries have orbital decay times short compared to their stellar evolution timescale, and are being observed in a short-lived phase just before merger. Moreover, the tail of the semi-major axis distribution at $a<a_{\rm crit}$ should be accompanied by a much larger number $\propto \dot{a}^{-1} \propto a^{9.5}$ at larger separation, where $\beta=9.5$ for dynamical tides. For example, if $\sim 10$ systems were found at $a/a_{\rm crit} = 0.5$, this should be accompanied by $\sim 10 \times 2^{9.5} \simeq 7000$ systems at $a \simeq a_{\rm crit}$. The large reservoir of systems with slow orbital decay would then feed a small tail of systems with rapid orbital decay at smaller separation.

\section{Conclusion}
\label{sec:conclusion}


Motivated by current and future surveys that find binaries with SGB or RGB primaries and stellar or substellar secondaries, MESA models for primary stars of mass $M_1=1, 2$ and $3\, M_\odot$ have been used to compute dynamical and equilibrium tidal dissipation rates. The resultant orbital decay rate was used to compare merger times for different primary and secondary masses and orbital separations, as well as different prescriptions for turbulent viscosity. The role of synchronization of the primary's spin and the Darwin instability have been taken into account. 


The dynamical tide dominates for close-in systems, with less evolved primaries. The equilibrium tide dominates for wider systems and more evolved primaries. The dividing line between the two depends on primary and secondary masses, as well as the prescription for reduced viscosity.

The tidal evolution depends sensitively on the primary star's mass. For stars of mass $M_1 \ga 1.3\, M_\odot$, equilibrium and dynamical tidal friction is strongly suppressed on the MS, and turns on suddenly during the sub-giant branch phase as the convective envelope deepens. For close-in systems this may result in orbital decay that proceeds rapidly compared to the stellar evolutionary timescale. The reason why the equilibrium tide is suppressed at the MS stage is because the energy dissipation rate depends on the convective envelope mass and the eddy velocity. These two physical quantities are much larger in the RGB phase. In the MS phase, the dynamical tide is suppressed because the wave can't propagate inward, and the convective envelope for $M_1 \ga 1.3\, M_\odot$ is very thin.

For $M_2=1$ - $10\, M_{\rm Jup}$, the low mass secondary cannot provide enough angular momentum to synchronize the spin. For synchronization to occur, $M_2$ must be on the order of $0.1M_\odot$. After it reaches the Darwin instability, the system merges soon after. Neither small ($M_2\lesssim 10 M_{\rm Jup}$) nor large ($M_2\gtrsim 100 M_{\rm Jup}$) companions can give rise to fast orbital decay.  Small mass companions exert weak tidal forces, thereby causing low energy dissipation rates. High mass companions synchronize their orbits quickly, resulting in a small forcing frequency and consequently low energy dissipation rate. Only intermediate mass ($10\lesssim M_2 \lesssim 100 M_{\rm Jup}$) secondaries can sustain a large energy dissipation rate.

By accounting for the Darwin instability, dynamical tide, and Zahn's prescription for the reduced viscosity for the equilibrium tide, we define a critical separation $a_{\rm crit}$ below which the system merges rapidly compared to the stellar evolution timescale.  Figures  \ref{acrit_1Msun.pdf}, \ref{acrit_2Msun.pdf} and \ref{acrit_3Msun.pdf} show  that the majority of APOGEE binaries show separations larger than $a_{\rm crit}$ for the observed $\log g$, indicating that these systems likely have very slow orbital decay. A small number of systems have observed separations $a<a_{\rm crit}$, implying rapid orbital decay. 

To compare to exoplanets and their host stars from the compilation in \citet{2014PASP..126..827H}, $a_{\rm crit}$ was computed over a finer grid of primary mass $M_1=1.0-1.5\, M_\odot$. The size of the convective core during the MS changes rapidly over this stellar mass range. As a result, the dynamical tide may turn on suddenly when the convective core disappears, and $a_{\rm crit}$ rapidly moves outward. Figure \ref{acrit_krxir_1015Msun_exo.pdf} compares our calculations to the exoplanet data. In a detailed comparison of the measured mass to the appropriate theory curve, we find that only a few systems have $a \la a_{\rm crit}$. However, this result must be taken with caution given the uncertainties in measuring primary gravity and mass. 


The comparison between the calculations of $a_{\rm crit}$ and semi-major axis data in this paper is an extension of \citet{2017MNRAS.470.2054C}, who studied hot Jupiters undergoing orbital decay due to the dynamical and equilibrium tides. \citet{2017MNRAS.470.2054C} computed orbital decay rates for five systems with detections or upper limits. These authors also found that the dynamical tide dominated for short-period orbits, and that radiative diffusion damping is important during the post main sequence phase. The present paper is focused on systems with no measured orbital decay rate, for which the only comparison is with the distribution of semi-major axis. Our study extends \citet{2017MNRAS.470.2054C} by carrying out a parameter study for a wide range of stellar mass, evolutionary state of the star, and companion mass. Our study also includes a criterion for the traveling wave limit to apply due to either strong radiative diffusion damping or nonlinear wave breaking when $k_r \xi_r \ga 1$ at the center.

\citet{2013ApJ...772..143S} discussed observational evidence that SGBs with planetary-mass companions show a strong deficit of systems with $a \la 0.67\, \rm AU$ ($P_{\rm orb}=200$ days). Further, the closest systems at $0.67\, \rm AU$ had fairly circular orbits. They proposed a scenario to explain this with tides, in which tidal friction is weak on the MS but increases dramatically on the sub-giant branch. Their scenario requires $R_1 \sim 3$ - $4\, R_\odot$ primary stars to cause orbital decay out to $0.67\, \rm AU$, and circularization of the orbits just outside this. Our results show that, with the tidal friction mechanisms included here, that orbital decay rate can only affect systems out to $a=0.05$ - $0.15\, \rm AU$ for this range of radii (and depending on the stellar mass), and that tidal friction is many orders of magnitude too weak to affect systems at $a=0.67\, \rm AU$.


	
	


\section*{Acknowledgements}

MS thanks Chenliang Huang, Dom Pesce and Scott Suriano for the very useful suggestions in improving the paper. The authors thank the referee for comments which significantly improved the paper. This work was supported by NASA ATP grant NNX14AB40G.

\appendix

\section{Analytic Estimate of the Heating Rate}
\label{sec:analytic_eq_tide}

Analytic estimates for the dissipation rate can be derived by treating the convective envelope as an $n=3/2$ polytrope with interior mass $m(r) \simeq M_1$. The latter assumption greatly simplifies the formulae for the density profile, however it leads to factor of a few errors for un-reduced viscosity since the dissipation occurs sufficiently deep in the convection zone that the interior mass $m(r)$ is changing rapidly there (see the middle panel of Figure \ref{fig:viscosity_energy_dissipation_logP}). The approximation is better for linear and quadratic scalings, as their main contribution is closer to the surface. 

For $\ell=2$, the dissipation integral can be simplified to 
\begin{eqnarray}
\begin{split}
\dot{E}  \simeq & \frac{2016\pi}{5} (n-\Omega)^2 R_1^2 \left( \frac{M_2}{M_1} \right)^2 \left( \frac{R_1}{a} \right)^6 \\
&\times \int_{r_{\rm bcz}}^{R_1} dr \rho \nu \left( \frac{r}{R_1} \right)^8.
\end{split}
\end{eqnarray}
MESA models of a $M_1=1.5\, M_\odot$ RGB envelope gives a polytropic constant, $K=P/ \rho^{5/3}$, strongly dependent on the evolutionary phase as
\begin{eqnarray}
K & \simeq & 3.98 \times 10^{14}\, {\rm cm^4\, g^{-2/3}\, s^{-2}}\ \left( \frac{R_1}{R_\odot} \right)^{0.94}.
\end{eqnarray}
Integrating the equation of hydrostatic balance then gives the density profile
\begin{eqnarray}
\rho(r) & \simeq & 2.65\, {\rm g\, cm^{-3}}\, \left( \frac{M_1}{M_\odot} \right)^{3/2} \left( \frac{R_\odot}{R_1} \right)^{2.9}
\left( \frac{R_1-r}{r} \right)^{3/2}.
\end{eqnarray}
The pressure scale height is
\be
H &= & \frac{P}{\rho g} = \frac{2}{5} \frac{r(R_1-r)}{R_1}
\ee
which has a peak of $H=R_1/10$ at $r=R_1/2$.
\citet{1995A&A...296..709V} found the luminosity-radius relation is roughly $L_1/L_\odot \simeq (R_1/R_\odot)^{1.6}$.
The mixing length velocity $v_{\rm ed} \simeq (L_1/4\pi r^2 \rho)^{1/3}$ can then be used to give the ``standard" (non-reduced) viscosity

\begin{eqnarray}
\begin{split}
\nu_{\rm std} = & \frac{1}{3} v_{\rm ed} \alpha_{\rm mlt} H \simeq 5.34 \times 10^{13}\, {\rm cm^2\, s^{-1}}\, \left( \frac{M_\odot}{M_1} \right)^{1/2} \\
&\times \left( \frac{R_1}{R_\odot} \right)^{1.83}
 \left( \frac{r}{R} \right)^{5/6} \left( \frac{R_1-r}{R_1} \right)^{1/2},
\end{split}
\end{eqnarray}

which has a maximum inside the convection zone. The eddy turnover time is

\begin{eqnarray}
\begin{split}
\tau_{\rm ed} = \frac{\alpha_{\rm mlt} H}{v_{\rm ed}} \simeq & 346\, {\rm days}\ \left( \frac{M_1}{M_\odot} \right)^{1/2} \left( \frac{R_1}{10R_\odot} \right)^{0.17}\\
&\times \left( \frac{r}{R_1} \right)^{7/6} \left( \frac{R_1-r}{R_1} \right)^{3/2},
\end{split}
\end{eqnarray}
with peak value longer than the orbital periods of many APOGEE binaries with $P_f = P_{\rm orb}/2 \sim \rm days - weeks$. 

Given the run over these quantities with radius, the integrals for each viscosity model can now be performed. For un-reduced viscosity
the dissipation rate is
\begin{eqnarray}
\begin{split}
\dot{E}_{\rm std} = & 5.66 \times 10^{28}\, {\rm erg\ s^{-1}}\ \left( 1 - \Omega/n \right)^2 \left( \frac{M_\odot}{M_1} \right) \left( \frac{M_2}{M_\odot} \right)^2 \\
&\times \left( \frac{R_1}{R_\odot} \right)^{7.9}
\left( \frac{0.1\, \rm AU}{a} \right)^9,
\end{split}
\label{eq:Edotstd}
\end{eqnarray}
by comparison, the dissipation rate for the ``Zahn" (linear in $P_f$) turbulent viscosity is
\begin{eqnarray}
\begin{split}
\frac{ \dot{E}_{\rm Z} }{ \dot{E}_{\rm std} } = & 0.47\, \left( \frac{a}{0.1\, \rm AU} \right)^{3/2} \left( \frac{M_1}{M_\odot} \right)^{3/2} \left( \frac{M_1+M_2}{M_\odot} \right)^{1/2}   \\
&\times \left( \frac{R_\odot}{R_1} \right)^{0.19} \left( 1 - \frac{\Omega}{n} \right)^{-1},
\label{eq:EdotZ}
\end{split}
\end{eqnarray}
and the GN 
(quadratic in $P_f$) rate is
\begin{eqnarray}
\begin{split}
\frac{ \dot{E}_{\rm GN} }{ \dot{E}_{\rm std} } = & 0.027\, \ln \Lambda \left( \frac{a}{0.1\, \rm AU} \right)^{3} \left( \frac{M_1}{M_\odot} \right) 
 \left( \frac{R_\odot}{R_1} \right)^{0.34}\\
& \times \left( 1 - \frac{\Omega}{n} \right)^{-2}.
\label{eq:EdotGN}
\end{split}
\end{eqnarray}
Here $\ln \Lambda \equiv \int_{P_{\rm min}}^{P_{\rm max}} d\ln P \simeq {\rm a\ few} $ represents the flat integrand observed for the GN curve in Figure  \ref{fig:viscosity_energy_dissipation_logP}.

The Zahn and GN scalings are shallower with orbital separation and stellar radius, and have a different dependence on forcing frequency. For nearly synchronous rotation, the forcing frequency $2(n-\Omega)$ becomes small, and un-reduced viscosity is appropriate.

Equations \ref{eq:EdotZ} and \ref{eq:EdotGN} can be set to unity and solved for the critical semi-major axis inside of which reduced viscosity operates. Zahn's prescription holds for
\be
\begin{split}
a \leq & 0.17\, {\rm AU} \left( \frac{M_\odot}{M_1} \right) \left( \frac{M_\odot}{M_1+M_2} \right)^{1/3} \left( \frac{R_1}{R_\odot} \right)^{0.11}\\
& \times \left( 1 - \frac{\Omega}{n} \right)^{2/3}
\end{split}
\label{eq:reduced_criterion}
\ee
and a similar expression holds for the GN prescription, with a slightly different numerical coefficient reflecting the coefficients $1/2$ and $1/(2\pi)^2$ in two prescriptions.

\section{Analytic Estimate of Critical Semi-major Axis}
\label{sec:analytic_acrit}

For binaries with high mass companions, the spin of the primary star will synchronize to the orbit, and orbital decay then proceeds on the stellar evolution timescale. This long phase of evolution ends when 
\be
\begin{split}
	a \leq  a_D \simeq 6\, R_1 \left( \frac{I_1}{0.12\, M_1 R_1^2} \right)^{1/2} \times \left( \frac{0.01\, M_1}{M_2} \right)^{1/2},
	\label{eq:aD_vs_R}
\end{split}
\ee
as the orbital decay will accelerate and the rotation rate of the primary will no longer be synchronized. An analytic calculation of $a_{\rm crit}$ in the synchronized case is complicated. The use of $a_{\rm D}$ as $a_{\rm crit}$ is not a good approximation as typically there has been orbital decay before the instability is reached, and also because there some further expansion of the primary after $a=a_{\rm D}$. Our numerical results show that $a_{\rm D}$ is typically smaller than $a_{\rm crit}$ by a factor of $\sim 2$.

Next, analytic scalings for $a_{\rm crit}$ are derived for the equilibrium tide, assuming $\Omega \ll n$ and $M_2\ll M_1$. Plugging the viscous heating rate (Equation \ref{eq:Edotstd} to \ref{eq:EdotGN}) into Equation \ref{dota_dotE} gives the following orbital decay rates for each viscosity formula to be
\be
\begin{split}
	\dot{a}_{\rm{STD}} = & - 9.65\times 10^{-7} \,{\rm cm \, s^{-1}}\,  \bigg(\frac{M_2}{M_1}\bigg) \bigg( \frac{R_1}{R_{\odot}}\bigg)^{7.93} \\
	& \times  \bigg( \frac{a}{0.1 \rm AU}\bigg)^{-7},
	\label{adotSTD}
\end{split}
\ee

\be
\begin{split}
	\dot{a}_{\rm{Z}} = & - 4.57\times 10^{-7} \,{\rm cm \, s^{-1}}\, \left( \frac{M_1}{M_\odot} \right)^{-2} \bigg(\frac{M_2}{M_\odot}\bigg)\nonumber \\ 
	& \times \bigg( \frac{R_1}{R_{\odot}}\bigg)^{7.74} \bigg( \frac{a}{0.1 \rm AU}\bigg)^{-11/2},
	\label{adotZ}
\end{split}
\ee

and

\be
\begin{split}
	\dot{a}_{\rm{GN}} = & - 2.7\times 10^{-8} \,{\rm cm \, s^{-1}}\, \left( \frac{M_1}{M_\odot} \right)^{-3} \bigg(\frac{M_2}{M_\odot}\bigg)\\ 
	& \times \bigg( \frac{R_1}{R_{\odot}}\bigg)^{7.56} \bigg( \frac{a}{0.1 \rm AU}\bigg)^{-4}.
	\label{adotGN}
\end{split}
\ee

Each formula has the form $\dot{a}=-f(t) a^{-\beta}$ where the time-dependence has been parametrized in terms of stellar radius here. The critical semi-major axis for which the orbit can decay to $a=0$ in a time $t$ is
\be
a_{\rm crit}(t) & = &\left[ (\beta+1) \tau(t) \right]^{1/(\beta+1)}
\ee
where
\be
\tau(t) &= & \int_0^t dt' f(t')
\ee
is a new time coordinate with the units $(\rm length)^{\beta+1}$. 

The time integral may be simply performed for RGB stars \citep{1995A&A...296..709V}. Since the radius and shell-burning luminosity mainly depend on the helium core mass, $M_{\rm He,1}$, a change of variables from $t$ to $R_1$ may be found using
\be
\dot{R}_1 & = & \frac{dR_1}{dM_{\rm He,1}}\frac{ dM_{\rm He,1} }{dt} 
\simeq  \left( \frac{R_\odot}{3.7\, \rm Gyr} \right) \left( \frac{R_1}{R_\odot} \right)^{2.3}.
\label{eq:Rdot}
\ee
The time integrals can then be written
\be
\tau(t) & \simeq & 3.7\, {\rm Gyr}\, \int_0^{R_1/R_\odot} dx f(t) x^{-2.3},
\label{relation_t_R}
\ee
where $x=R_1/R_\odot$. Since $f(t)$ has been expressed as a power of $R_1$, the integrals can be directly evaluated, and are dominated by the largest $x$. The results for each viscosity formula are then

\be
a_{\rm crit,STD}(t) \simeq 0.37\, {\rm AU}\, \left( \frac{M_2}{0.1\, M_1} \right)^{1/8} \left( \frac{R_1}{10\, R_\odot} \right)^{0.83} 
\label{eq:acritstd}
\ee

\be
a_{\rm crit,Z}(t) \simeq 0.41\, {\rm AU}\, \left( \frac{M_\odot}{M_1} \right)^{4/13}
\left( \frac{M_2}{0.1\, M_\odot} \right)^{2/13}\, \left( \frac{R_1}{10\, R_\odot} \right)^{0.99} 
\label{eq:acritz}
\ee

\be
a_{\rm crit,GN}(t) \simeq 0.35\, {\rm AU}\, \left( \frac{M_\odot}{M_1} \right)^{3/5}
\left( \frac{M_2}{0.1\, M_\odot} \right)^{1/5}\, \left( \frac{R_1}{10\, R_\odot} \right)^{1.25}
\label{eq:acritgn}
\ee
while each expression has a similar value for these fiducial parameters, their scalings with $M_1$, $M_2$ and $R_1$ differ. As $M_2 \ll M_1$, $\xi_{r,\ell m} \sim \xi_{h,\ell m}$ and $\Omega \ll n$ are assumed, there is a disagreement between the factors before the scalings from Equation \ref{adotSTD} to \ref{eq:acritgn} and the numerical result. The factors in Equation \ref{eq:acritz} are 0.28, 0.26 and 0.25 AU for $M_1=1,\,2$ and $3M_{\odot}$, respectively. The main purpose of showing these equations is to find how $a_{\rm crit}$ scales with $M_1$, $M_2$ and $R_1$.

The critical semi-major axis for reduced viscosity is only relevant if $\tau_{\rm ed} \ga P_f$ at the critical radius. Plugging Equation \ref{eq:acritz} into Equation \ref{eq:reduced_criterion} shows that the Zahn prescription applies for
\be
R_1 & \leq & 5.1\, R_\odot \, \left( \frac{0.1\, M_\odot}{M_2} \right)^{0.18} \left( \frac{M_1}{M_\odot} \right)^{1.31} \ \ \ {\rm (Zahn)}
\\
R_1 & \leq & 10\, R_\odot\, \left( \frac{0.1\, M_\odot}{M_2}\right)^{0.26}\left( \frac{M_1}{M_\odot} \right)^{1.1} \ \ \ {\rm (GN)}.
\ee
So if a particular system has $a \la a_{\rm crit}(t)$ during the time when $R_1$ is less than these critical values, then reduced viscosity should be used
rather than un-reduced viscosity. Further up the giant branch the un-reduced viscosity would apply.

Next an approximate expression is derived for $a_{\rm crit}$ for the dynamical tide. 
In Section \ref{Dynamical Tide Dissipation Rate} it was found that at fixed semi-major axis,  $L_{\rm dyn}$ increased strongly on the SGB and was nearly constant on the RGB. The numerical results for the RGB can be fit with the form

\be
\label{L_dyn_1Msun}
L_{\rm dyn}=C_{L,\rm dyn}\bigg(\frac{M_1+M_2}{M_{\odot}}\bigg)^{11/6}\bigg(\frac{M_2}{M_{\odot}}\bigg)^2
\bigg(\frac{a}{0.1{\rm AU}}\bigg)^{-23/2}
\label{L_dyn_general}
\ee
where $C_{L,\rm dyn}= 3.51\times10^{30}, 7.02\times10^{31}$, and $3.36\times10^{35}{\rm erg\,s^{-1}}$ for $M_1=1,\,\,2$ and $3M_{\odot}$, respectively. Plugging Equation \ref{L_dyn_general} into Equation \ref{dota_dotE}, and using Equation \ref{eq:Rdot} to convert age to stellar radius gives the final result

\be
\begin{split}
	a_{\rm crit}= & C_{a_{\rm crit},\rm dyn}\bigg(\frac{M_1+M_2}{M_{\odot}}\bigg)^{11/63}\bigg(\frac{M_2}{M_1}\bigg)^{2/21} \\
	& \times \bigg[  \bigg(\frac{R_{\rm brgb}}{R_{\odot}}\bigg)^{-1.3} - \bigg(\frac{R_1}{R_{\odot}}\bigg)^{-1.3} \bigg]^{2/21}
\end{split}
\label{acrit_dyn_R}
\ee
the coefficient has the value $C_{a_{\rm crit}}=$ 0.14, 0.19 and 0.42 AU for $M_1=1,\,\,2$ and $3M_{\odot}$, respectively. 
Unlike the equilibrium tide, the integral over time is dominated by the base of the RGB for the dynamical tide, and a lower limit $R_{\rm brgb}$ has been assumed for the radius there. Hence $a_{\rm crit}$ asymptotes to a constant as $R_1$ grows, allowing the equilibrium tide to dominate for wide orbits. Similar to the scaling functions for the equilibrium tide, Equation \ref{L_dyn_general} and \ref{acrit_dyn_R} is qualitatively right with the scalings of $M_1$, $M_2$ and $R_1$, but they are not in a good agreement with the numerical result for the entire SGB and RGB phase. Because the wave luminosity is not a constant in SGB phase, and the equilibrium tide is the main mechanism for orbital shrinking in RGB phase.

All the results presented in Section \ref{sec:examples} used sufficiently large $M_2$ that $k_r\xi_r > 1$ in the core, giving rise to traveling waves. 
The dynamical tide due to smaller, planetary mass companions may still generate the traveling wave limit of the dynamical tide if $P_{\rm orb} > P_{\rm orb, diff}$ and radiative diffusion damping is strong. However, if $M_2$ is too small, the orbit will not decay, but rather the star will expand out to meet the planet. The lower limit to $M_2$ that has $a_{\rm crit} > R_1$ may be estimated from Equation \ref{acrit_dyn_R}. For $M_1=1\, M_\odot$, in the limit $R_1 \ga R_{\rm brgb}$, for simplicity, the result is
\be
M_{\rm 2,min} & \simeq & 3\, M_\oplus\ \left( \frac{R_1}{10\, R_\odot} \right)^{21/2} \left( \frac{R_{\rm brgb}}{5\, R_\odot} \right)^{1.3}.
\ee
Hence there is a small parameter space for sub-Jupiter-sized planets to have a modest amount of orbital decay prior to the merger.

\end{document}